\documentclass[nohyper]{JHEP3}
\usepackage{cite}
\usepackage{epsfig}
\usepackage{amsmath}

\newcommand{\mycomm}[1]{\hfill\break $\phantom{a}$\kern-3.5em{\tt===$>$ \bf #1}\hfill\break}
\newcommand{\mycommA}[1]{\hfill\break $\phantom{a}$\kern-3.5em{\tt   $>$ \bf #1}\hfill\break}

\newcommand{\be}{\begin{equation}}
\newcommand{\ee}{\end{equation}}
\newcommand{\ba}{\begin{eqnarray}}
\newcommand{\ea}{\end{eqnarray}}

\def\tHooft{\hbox{\tiny 't Hooft}}
\def\eq#1{Eq.~(\ref{#1})}

\def\MSbar{\hbox{\tiny ${\overline{\rm MS}}$}}

\def\sing{\hbox{\tiny sing.}}
\def\reg{\hbox{\tiny reg.}}

\def\SDG{\hbox{\tiny SDG}}

\def\LO{\,{\rm LO}\, }

\catcode`\@=11 
\def\lsim{\mathrel{\mathpalette\@versim<}}
\def\gsim{\mathrel{\mathpalette\@versim>}}
\def\@versim#1#2{\vcenter{\offinterlineskip
        \ialign{$\m@th#1\hfil##\hfil$\crcr#2\crcr\sim\crcr } }}
\catcode`\@=12 

\title{Running--coupling effects in the
triple--differential
charmless semileptonic decay width}

\author{Paolo Gambino\\
\\
INFN, Sezione di Torino, and Dipartimento di Fisica Teorica,\\
Universit\`a di Torino, Via P. Giuria 1, I-10125 Torino, Italy }

\author{Einan Gardi \\
Cavendish Laboratory, University of Cambridge\\
Madingley Road, Cambridge, CB3 0HE, UK,\\
and\\
Department of Applied Mathematics \& Theoretical Physics,\\
Wilberforce Road, Cambridge CB3 0WA,~UK
}

\author{
Giovanni Ridolfi\\
INFN, Sezione di Genova, and Dipartimento di Fisica,\\
Universit\`a di Genova, Via Dodecaneso 33, I-16146 Genova,  Italy }

\abstract{We compute the fully--differential ${\bar B}\to X_u l \bar{\nu}$ decay width
 to all orders in perturbation theory in the large--$\beta_0$ limit.
Each of the five structure functions that build the hadronic tensor
is expressed as a Borel integral, summing up ${\cal O} (C_F
\beta_0^{n-1} \alpha_s^n)$ corrections for any $n$. We derive
analytic expressions for the Borel transforms of both real and
virtual diagrams with a single dressed gluon,
and perform an all--order infrared subtraction,
where the Borel parameter serves also as an infrared regulator.
Expanding the result we recover the known
triple--differential NLO coefficient, and obtain an explicit
expression for the ${\cal O} (C_F \beta_0 \alpha_s^2)$
triple--differential NNLO correction. This result can be used to
improve the determination of $|V_{\rm ub}|$ from inclusive ${\bar
B}\to X_u l \bar{\nu}$ measurements at the B factories with a
variety of kinematic cuts. }

\keywords{inclusive B decay, resummation, renormalons, heavy quarks}

\preprint{Cavendish-HEP-06/22\\
GEF-TH-16/2006\\
DFTT-25/2006}

\begin{document}

\section{Introduction}

The measurements~\cite{unknown:2006bi,HFAG} of semileptonic $b$ decay
branching fractions (BF) play a crucial r\^ole in the determination of
the CKM matrix elements~\cite{Charles:2004jd,Bona:2005vz}, which form
the basis for many precision tests of the Standard Model and provide
an input for new physics searches.  While any potential discovery of
new physics in the flavor sector is associated with loop--induced
transitions, the CKM parameters are most reliably determined by
tree--level weak decays.  Here two fundamental ingredients are
$|V_{\rm cb}|$ and $|V_{\rm ub}|$, which are measured in semileptonic
$b\to c$ and $b\to u$ decays, respectively.

Both inclusive and exclusive semileptonic measurements are used to
extract these parameters. Inclusive measurements are inherently more
robust owing to their limited sensitivity to the hadronic structure of
the initial and final states. However, since $b\to u$ transitions are
about 50 times less abundant than $b \to c$ ones, kinematic cuts must
be applied in order to isolate the $b \to u$ decays and measure
$|V_{\rm ub}|$. Consequently, the calculation of the fully
differential spectrum is essential for precision measurements of
$|V_{\rm ub}|$.

The theoretical calculation of inclusive decay spectra is complicated
by the presence of large perturbative and non-perturbative
corrections.  In $b$ decays into light quarks, e.g. $\bar{B}\to X_s
\gamma$ and $\bar{B}\to X_u l \bar{\nu}$, most events are
characterized by jet--like momentum configurations, where the
invariant mass of the hadronic system in the final state is small.
When computing the differential spectrum, or the BF with kinematic
cuts, one encounters parametrically--large Sudakov logarithms as well
as non-perturbative corrections that are associated with the momentum
distribution of the $b$ quark in the
meson~\cite{Neubert:1993um,Neubert:1993ch,Bigi:1993ex,Korchemsky:1994jb,Falk:1993vb,Lange:2005yw,Lange:2005qn,Aglietti:2006yb,Aglietti:2005mb,Aglietti:2001br}.

Recently, there has been significant progress in the application of
resummed perturbation theory to compute inclusive decay spectra
using the method of Dressed Gluon Exponentiation
(DGE)~\cite{Andersen:2005bj,Andersen:2005mj,Andersen:2006hr,Gardi:2006gt,Gardi:2006jc}.
Underlying this approach is the realization that
\emph{running--coupling corrections} play an important r\^ole in
shaping the spectrum. Beyond the purely perturbative issue, infrared
renormalons are useful in consistently separating between
perturbative and non-perturbative corrections while retaining the
predictive power of perturbation theory.

The significance of running--coupling corrections stems from the fact that the
gluon virtuality, which sets the \emph{effective scale} of the
coupling~\cite{Brodsky:1982gc},
is typically lower, sometimes significantly lower, than the
hard scale $m_b$ that is used as the default renormalization point.
Consider for example the fully--integrated $b\to X_u l \bar{\nu}$ width,
\begin{equation}
\Gamma(b\to X_u l \bar{\nu}) =\frac{G_F^2\,|V_{\rm ub}|^2\,m_b^5}{192\pi^3}
\left[1\,+\,b_1\frac{\alpha_s^{(N_f+1)}(m_b)}{\pi}
\,+\,b_2\left(\frac{\alpha_s^{(N_f+1)}(m_b)}{\pi}\right)^2+\cdots\right],
\label{total}
\end{equation}
which is known to
NNLO~\cite{vanRitbergen:1999gs},
\begin{align}
\label{b_n}
\begin{split}
b_1&=C_F \bigg(\frac{25}{8}-\frac{1}{2} \pi^2\bigg)\\
b_2&=\bigg(\frac{1009}{96}-\frac{77}{72} \pi^2-8 \zeta_3\bigg) C_F \beta_0\\
   &\hspace*{-10pt}+C_F^2 \bigg(\frac{11047}{2592}+\frac{53}{6} \pi^2 \ln(2)-
   \frac{515}{81} \pi^2-\frac{223}{36} \zeta_3
    +\frac{67}{720} \pi^4\bigg)
   \\&\hspace*{-10pt}+C_F\Bigg[\bigg(\frac{13759}{2592}+\frac{4061}{2592} \pi^2
   +\frac{145}{72} \zeta_3
    -\frac{53}{12} \pi^2 \ln(2)+\frac{101}{1440} \pi^4\bigg) C_A
   -\frac{85}{432} \pi^2+\frac{16987}{1152}-\frac{32}{3} \zeta_3\Bigg]\, ,
\end{split}
\end{align}
with
\begin{equation}
\label{beta0}
\beta_0=\frac{11}{12}C_A-\frac{1}{6} N_f\,,
\end{equation}
where $N_f$ is the number of \emph{light} flavors. Here we have
split the $b_2$ coefficient computed in
Ref.~\cite{vanRitbergen:1999gs} into a running--coupling part, proportional
to $\beta_0$, and the rest.
We find that with $N_f=4$ the former yields $b_2^{\beta_0}\simeq
-26.84$, while the latter $b_2^{\rm rest}\simeq 5.54$, adding up to
$b_2\simeq -21.30$. Evidently, the running--coupling corrections are
dominant. These large corrections are related to the leading infrared
renormalon of the pole mass $m_b$, which is located at $u=1/2$, where
$u$ is the relevant Borel parameter. The eventual
cancellation~\cite{Bigi:1994em,Beneke:1994sw,Beneke:1994bc} of this
ambiguity in \eq{total} involves the overall factor $m_b^5$ on the one
hand and the series in the square brackets on the other. This means,
in particular, that higher--order corrections ${\cal O} (C_F
\beta_0^{n-1} \alpha_s^n)$ in \eq{total} are large and form a
non-summable series. Owing to the proximity of the $u=1/2$ renormalon
to the origin of the Borel plane, and the relatively low scales
involved, the factorial divergence becomes relevant already at the
first few orders.

Let us consider now the \emph{differential} decay width. As usual,
the effective scale depends on the kinematics; in the region
selected by kinematic cuts, where the invariant mass of the hadronic
system is small, radiation is confined to be soft or collinear. An
obvious consequence is that the effective scale of the coupling gets
small, and therefore large running--coupling corrections should be
expected.

Moreover, the \emph{normalized} spectrum too is affected by infrared
renormalons. Despite the absence of the overall factor $m_b^5$,
infrared renormalons show up in the normalized spectrum because the
support of the on-shell decay width is set by $m_b$: an ${\cal
O}(\Lambda)$ variation of the pole mass amounts to an ${\cal
O}(\Lambda)$ \emph{shift} of the Born--level $\delta$--function
spectrum.  Therefore, \emph{all the moments} of the normalized $b$ decay
spectra, defined at the partonic level, have an infrared renormalon
ambiguity at $u=1/2$~\cite{Gardi:2004ia}.  In DGE this ambiguity,
which affects the Sudakov factor, gets cancelled against the pole mass
upon computing the resummed spectra in physical, hadronic
variables~\cite{Andersen:2005mj}. In other
approaches~\cite{Bosch:2004th,Gambino:2005tp} it is dealt with by
using an infrared cutoff and absorbing the soft contribution into the
definition of the non-perturbative parameters.  In any case, the
presence of this infrared sensitivity at the level of the partonic
calculation cannot be ignored. Computing decay spectra to higher
orders in perturbation theory, one therefore expects to find large
running--coupling corrections.

Recently, a first complete NNLO calculation of an inclusive decay
spectrum has been performed~\cite{Melnikov:2005bx,Asatrian:2006sm}
for the case of $\bar{B}\to X_s \gamma$ through the effective
magnetic dipole operator. A striking feature of this result is the
dominance of the ${\cal O}(C_F\beta_0\alpha_s^2)$ contribution
(which has been known since a while~\cite{Ligeti:1999ea}) with
respect to other color factors appearing at this order. The
similarity of the two processes, $\bar{B}\to X_s \gamma$ and
$\bar{B}\to X_u l \bar{\nu}$, and the dominance of running--coupling
corrections in the former, leave no doubt that these corrections are
dominant also in the latter.

In the case of the triple--differential $\bar{B}\to X_u l \bar{\nu}$
spectrum, the perturbative expansion is known in full to ${\cal
O}(\alpha_s)$ (NLO) only~\cite{DeFazio:1999sv}. NNLO corrections
have been computed in full only for the integrated width
\cite{vanRitbergen:1999gs}, \eq{total} above. In addition, ${\cal
O}(C_F\beta_0\alpha_s^2)$ real--emission corrections have been
recently computed~\cite{Hoang:2005pj} for one particular
single--differential spectrum, namely the distribution in the
(small) ``plus'' lightcone--momentum component,~$p_j^+$. The
complete ${\cal O}(C_F\beta_0^n\alpha_s^{n+1})$ have also been
computed numerically for the five structure functions of $\bar{B}\to X_c l
\bar{\nu}$ \cite{Aquila:2005hq}; the results for physical
observables obtained with a finite charm mass can in principle be
extrapolated numerically to the massless case, but this procedure
involves delicate numerical cancellations and lacks the flexibility
necessary in practical implementations.

Additional results beyond the NLO are available in the Sudakov
limit~\cite{Andersen:2005bj,Andersen:2005mj}: the Sudakov exponent has been
determined at NNLL accuracy~\cite{Korchemsky:1992xv,Gardi:2005yi,Moch:2004pa},
and to all--orders in the large--$\beta_0$ limit~\cite{Gardi:2004ia,Andersen:2005mj}.

In this paper we perform an all--order calculation
of the triple--differential $\bar{B}\to X_u l \bar{\nu}$
spectrum in the large--$\beta_0$ limit.
We derive analytic expressions for the Borel transform of
real and virtual diagrams with a single dressed gluon,
which represent the sum of
${\cal O} (C_F \beta_0^{n-1} \alpha_s^n)$ corrections for any $n$.
We then preform an all--order infrared subtraction directly in terms of the
Borel variable. By expanding the result we recover the known
triple--differential NLO coefficient~\cite{DeFazio:1999sv},
and obtain an explicit expression for the ${\cal O} (C_F \beta_0 \alpha_s^2)$
triple--differential NNLO correction. By integrating this expression we confirm
the results of the single--differential $p_j^+$ spectrum~\cite{Hoang:2005pj}
as well as the $\beta_0$ term in the integrated width\cite{vanRitbergen:1999gs}.

The ${\cal O} (C_F \beta_0 \alpha_s^2)$ triple differential width  we compute here
is an important ingredient in improving the determination of $|V_{\rm ub}|$
from inclusive ${\bar B}\to X_u l \bar{\nu}$ measurements with a variety of
kinematic cuts. In this paper we do not perform any numerical studies; these will be
reported on separately.

The structure of the paper is as follows. In
Sec.~\ref{sec:Definitions} we recall the kinematics and set up the
notation. Next, in Sec.~\ref{sec:real} we present the real--emission
``characteristic function'' based on the calculation of
Ref.~\cite{Aquila:2005hq}, which was performed with an off-shell
gluon. In Sec. \ref{sec:borel} we derive a Borel representation of
the real--emission corrections; in Sec.~\ref{sec:expand_borel} we
expand the Borel function to obtain explicit formulae for
coefficients at the first few orders. Next, in
Sec.~\ref{sec:Sudakov} we consider the Sudakov limit and extract the
non-integrable terms in the real--emission contribution to all
orders, confirming the results of~\cite{Gardi:2004ia}. Using the
Borel variable as an infrared regulator, we prepare the tools for an
all--order infrared subtraction. In Sec.~\ref{sec:Virt} we compute
the virtual diagrams, using a Borel--modified gluon propagator. We
then perform the subtraction of infrared singularities, directly in
terms of the Borel variable. In Sec.~\ref{sec:put_together} we
combine the results of both real and virtual diagrams for the
different structure functions, getting explicit expressions for the
coefficients of the triple--differential rate at NNLO. In
Sec.~\ref{sec:changing_variables} we demonstrate the way in which
infrared subtraction gets modified depending on the kinematic
variables used. Finally, in Sec.~\ref{sec:Conc} we summarize our
conclusions.

\section{Definitions and kinematics~\label{sec:Definitions}}

Let us write the triple--differential width of
\[
b(p)\longrightarrow l(p_l)+\bar{\nu}(p_{\bar {\nu}})+X_u(p_j)
\]
as
\begin{equation}
\label{triple_diff_light_cone}
\frac{d\Gamma}{dp_{j}^+\,dp_j^{-}\,dE_l}=\frac{G_F^2 \left|V_{\rm
ub}\right|^2}{16\pi^3}
\,L_{\mu\nu}(p_l,p_{\bar{\nu}}){\mathbb{W}}^{\mu\nu}(p,q),
\end{equation}
where the total momentum carried by the leptons is
$q=p_l+p_{\bar{\nu}}$ and $E_l$ is the energy of the charged lepton
in the $b$ rest frame. The momentum of the hadronic system is
expressed in terms of lightcone components, namely
\begin{equation}
p_j^{\pm}\equiv E_j\mp \left|\vec{p}_j\right|,
\end{equation}
so $p_j^{-}=\alpha m_b$ and $p_j^{+}=\beta m_b$ are the large and
small lightcone components of the ``jet'', respectively. They obey
\begin{equation}
\label{ps} 0\leq\beta\leq\alpha\leq 1.
\end{equation}
The relation of these variables with the invariant masses of the
``hadronic'' (partonic) and the leptonic systems, respectively, is
\begin{eqnarray}
\label{invariant_masses} p_j^2= m_b^2\alpha\beta\,;\qquad\qquad  q^2
= m_b^2(1-\alpha)(1-\beta),
\end{eqnarray}
where $0\leq p_j^2\leq \left(1-\sqrt{q^2}\right)^2$.
Throughout the paper, $m_b$ represents the pole mass of the $b$ quark.

It is convenient to write the normalization of the differential
width in terms of the total tree--level width
\begin{equation}
\label{Gamma0}
\Gamma_0\equiv \frac{G_F^2\left|V_{\rm
ub}\right|^2\,m_b^5}{192\pi^3},
\end{equation}
namely express \eq{triple_diff_light_cone} as:
\begin{equation}
\label{triple_diff_light_cone_alpha_beta_x}
\frac{d^3\Gamma}{d\alpha\,d\beta\,dx}= \frac12
m_b^3\,\frac{d\Gamma}{dp_{j}^+\,dp_j^{-}\,dE_l} = 6\,\Gamma_0\,
\frac{1}{m_b^2}\,
L_{\mu\nu}(p_l,p_{\bar{\nu}}){\mathbb{W}}^{\mu\nu}(p,q),
\end{equation}
where $x\equiv 2E_l/m_b$. Phase--space integration then yields:
\begin{eqnarray}
\label{phase_space} \Gamma_{b\to u}&=&\int_0^1d\alpha \int_0^1 dr
\alpha\int_{1-\alpha}^{1-r\alpha}dx\,\frac{d\Gamma}{d\alpha\,d\beta\,dx}\nonumber
\\&=& 6\Gamma_0\, \int_0^1 d\alpha \int_0^1 dr
\alpha\int_{1-\alpha}^{1-r\alpha}dx\,\frac{1}{m_b^2}\,
L_{\mu\nu}(p_l,p_{\bar{\nu}}){\mathbb{W}}^{\mu\nu}(p,q),
\end{eqnarray}
where we use $r\equiv \beta/\alpha$ following
\cite{Andersen:2005mj}.

The choice of the lightcone variables in \eq{triple_diff_light_cone}
is motivated by the fact that the final state is typically
jet--like: at Born level $p_j^2\equiv 0$ so most events are
characterized by $p_j^2\ll m_b^2$, namely $\beta\ll \alpha$. While
the first NLO calculation of the triple--differential $b\to
ul\bar{\nu}$ spectrum~\cite{DeFazio:1999sv} used other kinematic
variables, the advantages of lightcone variables have recently been
acknowledged by several
authors~\cite{Bosch:2004th,Hoang:2005pj,Aglietti:2005mb,Andersen:2005mj}.

For massless leptons, the leptonic tensor is given by:
\begin{equation}
\label{leptonic}
L^{\mu\nu}(p_l,p_{\bar{\nu}})=p_l^{\mu}p_{\bar{\nu}}^{\nu}
+p_l^{\nu}p_{\bar{\nu}}^{\mu}-p_l\cdot p_{\bar{\nu}}
g^{\mu\nu}-i\epsilon^{\mu\rho\nu\sigma}
{p_l}_{\rho}{p_{\bar{\nu}}}_{\sigma}.
\end{equation}
Lorentz decomposition of the hadronic tensor
${\mathbb{W}}^{\mu\nu}(p,q)$ gives rise to five scalar ``structure
functions'':
\begin{align}
\label{StructureFunctions}
\begin{split}
{\mathbb{W}}^{\mu\nu}(p, q) &=- {\mathbb{W}}_1(\alpha,\beta)
g^{\mu\nu} +{\mathbb{W}}_2(\alpha,\beta)v^{\mu}v^{\nu} +
i{\mathbb{W}}_3(\alpha,\beta)\epsilon^{\mu\nu\rho\sigma}v_{\rho}\hat{q}_{\sigma}\\
&+{\mathbb{W}}_4(\alpha,\beta)\hat{q}_{\mu} \hat{q}_{\nu}
+{\mathbb{W}}_5(\alpha,\beta) (v^{\mu}\hat{q}^{\nu} +
v^{\nu}\hat{q}^{\mu}),
\end{split}
\end{align}
where $v=p/m_b$ and $\hat{q}=q/m_b$. ${\mathbb{W}}^{\mu\nu}(p,q)$ is
related to $W^{\mu\nu}(p,q)$ defined in \cite{Aquila:2005hq} by:
\begin{align}
{\mathbb{W}}^{\mu\nu}(p,q)&= \frac{1}{\pi\,
m_b^2}\,{W}^{\mu\nu}(p,q)\left|
\frac{d(q^2,p_j^2)}{d(p_{j}^+,\,p_j^{-})}\right|=
\frac{(\alpha-\beta)}{\pi} {W}^{\mu\nu}(p,q) .
\end{align}
Note that both $W_i^{\mu\nu}(\alpha,\beta)$ and
${\mathbb{W}}_i^{\mu\nu}(\alpha,\beta)$  are dimensionless.
Contracting the Lorentz indices between the leptonic and hadronic
tensors, Eqs. (\ref{leptonic}) and  (\ref{StructureFunctions})
respectively, one finds:
\begin{align}
\label{WL}
\begin{split}
\frac{1}{m_b^2}L_{\mu\nu}(p_l,p_{\bar{\nu}}){\mathbb{W}}^{\mu\nu}(p,
q) \,=\,\,& (1-\alpha) (1-\beta)\,{\mathbb{W}}_1(\alpha,\beta)
 \\&-\frac12\bigg(x^2\,- \,x \, (2-\alpha-\beta)+(1-\alpha) (1-\beta)
 \bigg)\,
 {\mathbb{W}}_2(\alpha,\beta)\\&+
 (1-\alpha) (1-\beta) \left(x-1+\frac12 (\alpha+\beta)\right)\,
 {\mathbb{W}}_3(\alpha,\beta),
\end{split}
\end{align}
where, as above, $x=2E_l/m_b$. Each structure function
${\mathbb{W}}_i(\alpha,\beta)$ may be decomposed as:
\begin{equation}
\label{W_RV} {\mathbb{W}}_i(\alpha,\beta)=V_i(\alpha)\,\delta(\beta)
+R_i(\alpha,\beta).
\end{equation}
The functions $V_i(\alpha)$ and $R_i(\alpha,\beta)$ have
perturbative expansions in $\alpha_s(m_b)$. At the leading--order (LO),
$R_i(\alpha,\beta)=0$ and, for $i=1$ to 5,
\begin{eqnarray}
\label{LO} V_i^{\rm LO}(\alpha)=\left[
\alpha,\,4,\,2,\,0,\,-2\right].
\end{eqnarray}
Substituting (\ref{W_RV}) with (\ref{LO}) into (\ref{WL}) and using
(\ref{triple_diff_light_cone_alpha_beta_x}) one gets:
\begin{eqnarray}
\label{LO_result}
\frac{d^3 \Gamma^{\rm LO}}{d \alpha \,d\beta \,d x}=  \Gamma_0\,
\omega_0(\alpha, x)\,
\delta(\beta)\,; \qquad\qquad \omega_0(\alpha, x)\equiv 12\,(x+\alpha-1)\,(2-x-\alpha)
\end{eqnarray}
where $\Gamma_0$ is given in (\ref{Gamma0}).
As usual, this Born--level result receives perturbative corrections
from both virtual and real--emission diagrams. Purely virtual
contributions are proportional to $\delta(\beta)$. These, however,
contain infrared singularities that cancel against soft and
collinear real--emission singularities when performing phase--space
integration near the $\beta\to 0$ limit. Thus, beyond the LO, the
separate definition of $V_i(\alpha)$ and $R_i(\alpha,\beta)$ in
(\ref{W_RV}) requires a subtraction prescription. We shall return to
this issue in Sec.~\ref{sec:Sudakov}.

The full NLO, ${\cal O}(\alpha_s)$ result has been obtained in
Ref.~\cite{DeFazio:1999sv}, and checked in various papers that
considered higher--order running--coupling effects, including
Ref.~\cite{Gardi:2004ia} (real emission) and
Ref.~\cite{Aquila:2005hq} (real and virtual corrections). The result
was presented in terms of lightcone variables in
Ref.~\cite{Andersen:2005mj}.

\section{Real emission of an off-shell gluon and the characteristic function~\label{sec:real}}

Perturbative calculations of many observables in QCD, including
inclusive cross sections and decay rates can be improved by the
resummation of running--coupling
effects~\cite{Brodsky:1982gc,Brodsky:2000cr,Dokshitzer:1995qm,Beneke:2000kc,Beneke:1998ui,Ball:1995ni,Ball:1995wa,Aquila:2005hq}.
Specifically, keeping just the leading term in the $\beta$ function,
one sums up the terms proportional to $\beta_0^{n-1}\alpha_s^n$ to
all orders, the so called BLM terms.

Technically, running--coupling terms can be conveniently derived
using the dispersive method, see
e.g.~\cite{Dokshitzer:1995qm,Ball:1995ni,Ball:1995wa}, where the
one-loop calculation is performed with an off--shell gluon. The
calculation of the semileptonic decay ``structure functions''
$R_i(\alpha,\beta)$ with a single off-shell gluon was performed in
Ref.~\cite{Aquila:2005hq} for the more general case of $b\to c$
decay, where the charm mass $m_c$ is kept. Here we use this result
to derive the corresponding $b\to u$ result by sending $m_c\to 0$.
This limit leads to significant simplification that facilitates
obtaining closed form analytic expressions.

Further simplification is achieved by choosing the lightcone
variables described above, which are suitable for the final state
``jet'' kinematics. As we shall see, the result for
$R_i(\alpha,\beta)$ is completely symmetric under
$\alpha\longleftrightarrow \beta$; only the phase--space
restriction, \eq{ps}, breaks this symmetry.

The LO calculation of the real--emission diagrams with a
gluon of virtuality $m_g^2=\xi m_b^2$ yields:
\begin{eqnarray}
R_i(\alpha,\beta)\longrightarrow C_F\frac{\alpha_s(m_b)}{\pi}{\cal
F}_i(\alpha,\beta,\xi)+{\cal O}(\alpha_s^2)
\end{eqnarray}
with the following ``characteristic function'':
\begin{eqnarray}
\label{calF}
 {\cal
F}_i(\alpha,\beta,\xi)=\lambda^{-y_i}\left[\frac{1}{\left(\alpha\beta\right)^{z_i}}\,
\left(\frac{1}{\tau^-}-\frac{1}{\tau^+}\right)
\,P_i(\alpha,\beta,\xi)\,\,+\,\frac{1}{\alpha\beta}\,\ln\left(\frac{\tau^+}{\tau^-}\right)\,
Q_i(\alpha,\beta,\xi)\right]
\end{eqnarray}
 where the powers are $y_i=[1,2,1,2,2]$ and
$z_i=[3,3,3,2,3]$ for $i=1$ through 5, respectively, and
$P_i(\alpha,\beta,\xi)$ and $Q_i(\alpha,\beta,\xi)$ are polynomials
in all their arguments. Finally,
\begin{equation}
\label{tau_pm} \tau^\pm=
\xi\left(1-\frac{1}{2\alpha}-\frac{1}{2\beta}\right)
-\frac{1}{2}\left(\alpha+\beta\right)\pm\frac{1}{2}\sqrt{\lambda}\left(1-\frac{\xi}{\alpha
\beta}\right)
\end{equation}
where
\begin{equation}
\label{lambda_def} \sqrt{\lambda}=\alpha-\beta.
\end{equation}

Note that the phase--space limits are
\begin{equation}
0\leq \xi\leq \alpha\beta,
\end{equation}
where the upper limit corresponds to the situation where the entire
mass of the hadronic system $p_j^2=m_b^2\alpha\beta$ is given by the
gluon virtuality.

\section{Borel representation~\label{sec:borel}}

The Borel representation of the result for $R_i$ in the
single--dressed--gluon (SDG) approximation is:
\begin{eqnarray}
\label{Ri_borel} R_i^{\SDG}(\alpha,\beta)&=& \frac{C_F}{\beta_0}
\int_0^{\infty}du\,T(u)\,\left(\frac{\Lambda^2}{m_b^2}\right)^u\,B_i^{\SDG}(\alpha,\beta,u)\\
&=& C_F\,\bigg[c_i^{(1)}(\alpha,\beta)\,\frac{\alpha_s(m_b)}{\pi}+
c_i^{(2)}(\alpha,\beta)\,\beta_0\left(\frac{\alpha_s(m_b)}{\pi}\right)^2+\cdots\bigg]\,,
\nonumber
\end{eqnarray}
where $m_b$ is the bottom pole mass, $\Lambda$ and $\alpha_s$ are
defined in the $\overline{\rm MS}$ scheme, and $\beta_0$ is defined
in \eq{beta0}. $T(u)$ is the inverse Laplace transform of the coupling
(see Appendix A); in
the large--$\beta_0$ limit, where the renormalization--group
equation is just one loop, $T(u)\equiv 1$. Resummation of
running--coupling corrections beyond this strict limit can also be
performed using (\ref{Ri_borel}). This is briefly explained in
Appendix~\ref{sec:scheme_invariant_Borel}.

The Borel function in \eq{Ri_borel} can be derived from the
following integral over the characteristic function (see
e.g.~\cite{Ball:1995ni,Gardi:1999dq} or Sec. 2.2. in
\cite{Gardi:2006rp}),
\begin{eqnarray}
\label{B_u}
 B_i^{\SDG}(\alpha,\beta,u)&\equiv&-{\rm e}^{\frac53 u}\,\frac{\sin\pi u}{\pi
 }\,\int_0^{\alpha\beta}\frac{d\xi}{\xi}\,
 {\cal F}_i(\alpha,\beta,\xi)\,\xi^{-u},\nonumber\\
 &=&-{\rm e}^{\frac53 u}\,\frac{\sin\pi u}{\pi
 }\,\left(\alpha\beta\right)^{-u}b_i(\alpha,\beta,u),
\end{eqnarray}
where ${\cal F}_i(\alpha,\beta,\xi)$ is given in \eq{calF} and, upon changing variables
from $\xi$ to $\eta$ where $\xi\equiv (1-\eta)\alpha\beta$,
\begin{eqnarray}
\label{b_u} b_i(\alpha,\beta,u)\,\equiv\,
\int_0^{1}d\eta\,(1-\eta)^{-1-u}\, {\cal
F}_i(\alpha,\beta,(1-\eta)\alpha\beta).
\end{eqnarray}
We have performed the integral in (\ref{b_u}) analytically, and
checked the result by numerical evaluation. Below we give a few
details of the calculation and summarize the analytic expressions
for $b_i(\alpha,\beta,u)$.

Writing $\tau^{\pm}$ of \eq{tau_pm} in terms of $\eta$ we have
$\tau^{\pm}=(\alpha\beta-\alpha-\beta)(1-\kappa_{\pm}\eta)$, with
\begin{equation}
\label{kappa_def}
\kappa_{\pm}
\equiv\frac{\alpha\beta-\frac12(\alpha+\beta\pm\sqrt{\lambda} )}
{\alpha\beta-\alpha-\beta}
\end{equation}
so using \eq{lambda_def}
$\kappa_{+}=(\alpha\beta-\alpha)/(\alpha\beta-\alpha-\beta)$
 and $\kappa_{-}=(\alpha\beta-\beta)/(\alpha\beta-\alpha-\beta)$.
Given \eq{calF}, the basic integral needed in \eq{b_u} is of the
form
\begin{equation}
\label{basic_hyper} \int_0^1d\eta (1-\eta)^{-u}
\frac{1}{1-\kappa_{\pm}\eta}=\frac{1}{1-u} \,\, _2F_1\Big([1,
1],[2-u],\kappa_{\pm}\Big).
\end{equation}
All the terms in \eq{b_u} can be expressed in terms of this specific
hypergeometric function with the two assignments of the argument,
$\kappa_{\pm}$, and some additional rational functions. For example,
to integrate the log term in \eq{calF} times $(1-\eta)^j$ where $j$ is
a positive integer (to account for $Q_i(\alpha,\beta,\xi)$ that are
quadratic in $\xi$ we need $j=0,1,2$) one first integrates by parts
and then uses \eq{basic_hyper} to obtain:
\begin{align}
\label{ln_int}
\begin{split}
\int_0^1d\eta \,(1-\eta)^{-1-u+j}
\ln(1-\kappa_{\pm}\eta)&=\frac{-\kappa_{\pm}}{j-u}\,\, \int_0^1d\eta
\,(1-\eta)^{j-u} \frac{1}{1-\kappa_{\pm}\eta}\\&
=\frac{-\kappa_{\pm}}{(j-u)(j-u+1)} \,\, _2F_1\Big([1,
1],[2-u+j],\kappa_{\pm}\Big).
\end{split}
\end{align}
Finally, there are known identities that facilitate integer shifts of the
indices of hypergeometric functions. For example, to express the
hypergeometric function in~\eq{ln_int} in terms of our basic one in
\eq{basic_hyper} we need to shift the third index
from $[2-u+j]$ into $[2-u]$. This is straightforward to do
using~Eq. (2.10) in Ref.~\cite{Kalmykov:2006pu}.

The final result for $b_i(\alpha,\beta,u)$ takes the form
\begin{align}
\label{bi}
\begin{split}
 b_i(\alpha,\beta,u)=\lambda^{-y_i}
\Bigg[&{D_i(\alpha,\beta,u)} \bigg(\,\, _2F_1\Big([1,
1],[2-u],\kappa_{+}\Big) -\,\, _2F_1\Big([1,
1],[2-u],\kappa_{-}\Big)
\bigg)\\
&+ \sqrt{\lambda}\, S_i(\alpha,\beta,u) \bigg(\,\, _2F_1\Big([1,
1],[2-u],\kappa_{+}\Big) +\,\, _2F_1\Big([1,
1],[2-u],\kappa_{-}\Big)
\bigg)\\
&+ \sqrt{\lambda} \,T_i(\alpha,\beta,u)\Bigg],
\end{split}
\end{align}
where the entire $u$ dependence of the coefficient functions
is summarized by
\begin{align}
\label{DST}
\begin{split}
D_i(\alpha,\beta,u)&=
\frac{D_{i,0}(\alpha,\beta)}{u}+\frac{D_{i,1}(\alpha,\beta)}{1-u}
+\frac{\widetilde{D}_{i,1}(\alpha,\beta)}{(1-u)^2}
+\frac{D_{i,2}(\alpha,\beta)}{2-u}\\
S_i(\alpha,\beta,u)&=\frac{S_{i,0}(\alpha,\beta)}{u}+\frac{S_{i,1}(\alpha,\beta)}{1-u}
+\frac{\widetilde{S}_{i,1}(\alpha,\beta)}{(1-u)^2}
+\frac{S_{i,2}(\alpha,\beta)}{2-u}\\
T_i(\alpha,\beta,u)&=\frac{T_{i,0}(\alpha,\beta)}{u}+\frac{T_{i,1}(\alpha,\beta)}{1-u}
+\frac{T_{i,2}(\alpha,\beta)}{2-u}
\end{split}
\end{align}
and where $D_{i,j}(\alpha,\beta)$, $S_{i,j}(\alpha,\beta)$ and
$T_{i,j}(\alpha,\beta)$ are rational functions of $\alpha$ and $\beta$.
The explicit expressions are given
in Appendix \ref{sec:DST_explicit_expressions}.
We note that there are simple relations between some of
these functions.
For example, for any structure function $i$,
\begin{equation}
\label{D0S0_relation}
D_{i,0}(\alpha,\beta)=(\alpha+\beta-2\alpha\beta)\,S_{i,0}(\alpha,\beta).
\end{equation}

It is straightforward to check that there are no renormalon singularities
in $B_i^{\SDG}(\alpha,\beta,u)$ of \eq{B_u}. As usual, single poles in
$b_i(\alpha,\beta,u)$ are cancelled in $B_i^{\SDG}(\alpha,\beta,u)$
by the $\sin(\pi u)$ factor,
which is associated with the gluon momentum
being timelike. The double pole in $b_i(\alpha,\beta,u)$ at $u=1$,
which could have resulted in a single pole in $B_i^{\SDG}(\alpha,\beta,u)$,
is in fact not there: according to Eqs. (\ref{bi}) and (\ref{DST})
its residue is proportional to
\begin{equation}
(2-\kappa_{+}-\kappa_{-})\sqrt{\lambda}\widetilde{S}_{i,1}(\alpha,\beta)+
(\kappa_{+} - \kappa_{-})\widetilde{D}_{i,1}(\alpha,\beta)
\end{equation}
and therefore to $(\alpha+\beta)\widetilde{S}_{i,1}(\alpha,\beta)+
\widetilde{D}_{i,1}(\alpha,\beta)$, a combination that vanishes for
all $i$.

Thus, $R_i(\alpha,\beta)$ are free of infrared renormalons.
Nevertheless, the series for $R_i(\alpha,\beta)$ are divergent and they
are Borel summable only for large enough values
of $\beta$, owing to the convergence constraint on the Borel
integral in \eq{Ri_borel} for $u\to \infty$.
The consequences have been investigated in detail
in the context of the radiative decay~\cite{Andersen:2006hr}, see Sec.~2.3 there.
In moment space, the convergence constraint is replaced by infrared
renormalons~\cite{Gardi:2004ia} through the integration over $\beta$
near $\beta=0$, see \eq{Sud} below.

\section{Expanding the Borel function~\label{sec:expand_borel}}

In order to obtain the perturbative coefficients $c_i^{(n)} (\alpha,\beta)$
in the second line of
 \eq{Ri_borel} one expands the Borel function~$B_i^{\SDG}(\alpha,\beta,u)$ in powers of $u$, see
 \eq{expanding_Borel}.
The expansion of $_2F_1\Big([1, 1],[2-u],x\Big)$ is known, see
e.g.~Ref.~\cite{Kalmykov:2006pu}:
\begin{eqnarray}
\label{hyper_exp}
_2F_1\Big([1, 1],[2-u],x\Big)
&=&\frac{1-u}{x}\Bigg\{-\ln(1-x)+u\bigg[\frac{1}{2}\ln^2(1-x)+{\rm Li}_2(x)\bigg]\\ \nonumber &&\hspace*{-30pt}+u^2\left[-S_{1,2}(x)-\ln(1-x){\rm Li}_2(x)
+{\rm Li}_3(x)-\frac{1}{6}\ln^3(1-x)\right]\\ \nonumber &&\hspace*{-30pt}+u^3\bigg[-S_{2,2}(x)-\ln(1-x){\rm Li}_{3}(x)
+\ln(1-x)S_{1,2}(x)\\ \nonumber &&\hspace*{-10pt}+\frac{1}{2}\ln^2(1-x) {\rm Li}_2(x)+\frac{1}{24}\ln^4(1-x)
+S_{1,3}(x)+{\rm Li}_4(x)\bigg]+\cdots\Bigg\},
\end{eqnarray}
where Nielsen integrals are defined by
\begin{equation}
S_{a,b}(x)\equiv \frac{(-1)^{a+b-1}}{(a-1)!b!}
\int_0^1\frac{d\xi}{\xi}\ln^{a-1}(\xi)\ln^b(1-\xi x).
\end{equation}

Expanding $B_i^{\SDG}(\alpha,\beta,u)$ in \eq{B_u} in powers of $u$,
and using \eq{hyper_exp} to expand the hypergeometric functions, we
obtain the coefficients, expressed in terms of the functions of
Appendix \ref{sec:DST_explicit_expressions}. At ${\cal O}(\alpha_s)$,
using the definition of $\kappa_{\pm}$ in \eq{kappa_def} and the
relation of \eq{D0S0_relation}, we get:
\begin{align}
\label{c1}
\begin{split}
c_i^{(1)}(\alpha,\beta)&=
-\lambda^{-y_i}\Bigg\{-2(\alpha+\beta-\alpha\beta)
\ln\left(\frac{\beta}{\alpha}\right)\,\,S_{i,0}(\alpha,\beta)+\,
(\alpha-\beta)\,\,T_{i,0}(\alpha,\beta),
 \Bigg\}.
\end{split}
\end{align}
At ${\cal O}(\beta_0\alpha_s^2)$ we get
\begin{align}
\label{c2}
\begin{split}
c_i^{(2)}(\alpha,\beta)&=
-\lambda^{-y_i}\Bigg\{{D_{i,0}(\alpha,\beta)}
\bigg[\frac{ \frac{1}{2}\ln^2(1-\kappa_{+}) +{\rm
Li}_2(\kappa_{+})}{\kappa_{+}}-
\frac{ \frac{1}{2}\ln^2(1-\kappa_{-}) +{\rm
Li}_2(\kappa_{-})}{\kappa_{-}}\bigg]
\\&
\hspace{-8mm}
+\bigg(D_{i,1}(\alpha,\beta) -D_{i,0}(\alpha,\beta)
+\widetilde{D}_{i,1}(\alpha,\beta)
+\frac12D_{i,2}(\alpha,\beta)\bigg)
\bigg(\frac{\ln(1-\kappa_{+})}{-\kappa_{+}}-\frac{\ln(1-\kappa_{-})}{-\kappa_{-}}\bigg)\\
&\hspace{-8mm}
+\,{\sqrt{\lambda}}\,S_{i,0}(\alpha,\beta)
\bigg[\frac{ \frac{1}{2}\ln^2(1-\kappa_{+}) +{\rm
Li}_2(\kappa_{+})}{\kappa_{+}}+
\frac{ \frac{1}{2}\ln^2(1-\kappa_{-}) +{\rm
Li}_2(\kappa_{-})}{\kappa_{-}}
\bigg]\\
&\hspace{-8mm}
+\,{\sqrt{\lambda}}\,\bigg(S_{i,1}(\alpha,\beta)-S_{i,0}(\alpha,\beta)+\widetilde{S}_{i,1}(\alpha,\beta)
+\frac12S_{i,2}(\alpha,\beta)\bigg)
\left(\frac{\ln(1-\kappa_{+})}{-\kappa_{+}}+\frac{\ln(1-\kappa_{-})}{-\kappa_{-}}\right)\\
&\hspace{-8mm}
+{\sqrt{\lambda}}\left(T_{i,1}(\alpha,\beta)+\frac12T_{i,2}(\alpha,\beta)
\right)\Bigg\}
+\left(\frac{5}{3}-\ln(\alpha\beta)\right)\,c_i^{(1)}(\alpha,\beta).
\end{split}
\end{align}
Explicit expressions for $c_i^{(1,2)}(\alpha,\beta)$ that are
obtained upon substituting the functions in Appendix
\ref{sec:DST_explicit_expressions} and collecting terms, are listed
in Appendix \ref{sec:real_explicit}. In this way it is
straightforward to derive higher--order terms in the
single--dressed--gluon approximation.

\section{The Sudakov limit~\label{sec:Sudakov}}

Having derived an explicit result for the Borel function it is
straightforward to extract the singular terms in the $\beta\to 0$
limit, the Sudakov limit. The leading terms in this limit have
already been computed in Ref.~\cite{Gardi:2004ia}. They have been
extracted there --- see Sec.~3 and appendix B --- from an integral
representation of the real--emission result for the
triple--differential distribution, which was computed directly in
the Borel formulation; here we re-derive it from the Borel
representation (\ref{Ri_borel}) of the structure functions computed
in the previous section based on the dispersive method.

The Borel function is given in \eq{B_u} where $b_i(\alpha,\beta,u)$
are explicitly written in \eq{bi}. We wish to expand these results
at small $\beta$ keeping the other lightcone variable~$\alpha$ as
well as the Borel parameter $u$ fixed. In this limit
$\kappa_{-}=\beta(1-\alpha)/\alpha+{\cal O}(\beta^2)$ while
$\kappa_{+}=1-\beta/\alpha+{\cal O}(\beta^2)$. This means that in
\eq{b_u} $_2F_1\Big([1, 1],[2-u],\kappa_{-}\Big)$ can be readily
expanded at small $\beta$ while $_2F_1\Big([1,
1],[2-u],\kappa_{+}\Big)$ cannot. To extract the leading terms at
$\beta\to 0$ from the latter we first use the general identity:
\begin{equation}
\label{2F1_identity}
_2F_1\Big([1,1],[2-u],\kappa_{+}\Big)
=(1-u)\left[-\frac{1}{u}\, _2F_1\Big([1, 1],[1+u],1-\kappa_{+}
\Big)+\frac{\pi}{\sin\pi u}(1-\kappa_{+})^{-u}\kappa_{+}^{u-1}\right]
\end{equation}
The new hypergeometric function in
\eq{2F1_identity},
$_2F_1\Big([1, 1],[1+u],1-\kappa_{+})$, is of course expandable at
$\beta\to 0$, while the non-analytic contributions are explicitly
given by the second term in \eq{2F1_identity}.

With this replacement and using Eqs. (\ref{B_u}), (\ref{bi})
and (\ref{DST}) and the explicit expressions in
 Appendix \ref{sec:DST_explicit_expressions}, we obtain the
 expected singularity structure~\cite{Gardi:2004ia} (see Eq.~(3.17) in
 Ref.~\cite{Andersen:2005mj}) for small lightcone component $\beta$:
\begin{align}
\begin{split}
\label{B_i_small_beta}
 \left.B_{i}^{\SDG}(\alpha,\beta,u)\right\vert_{\beta\to 0}&= \frac{(\alpha\beta)^{-u}\,
 V_i^{\rm LO}(\alpha)}{\beta}
 \frac{ {\rm e}^{\frac53 u}}{u}\,\times \,\\ & \hspace*{-45pt}
\left[(1-u)\left(\frac{\beta}{\alpha}\right)^{-u}-
 \frac12\frac{\sin \pi u}{\pi u}\left(\frac{1}{1-u}+\frac{1}{1-u/2}
\right)
 \right]\,\times \Big(1+{\cal O}(\beta/\alpha)\Big),
\end{split}
\end{align}
where the last factor in \eq{B_i_small_beta}
serves as a reminder that integrable terms that are suppressed by powers of $\beta$
are excluded here.

As in the full $R_{i}(\alpha,\beta)$, there are additional ${\cal
O}(1/\beta_0)$ contributions to
$\left.B_{i}(\alpha,\beta,u)\right\vert_{\beta\to 0}$ starting at
${\cal O}(u^1)$, corresponding to ${\cal O}(\alpha_s^2)$, the NNLO.
These go beyond the large--$\beta_0$ limit, and therefore beyond the
calculation performed in the present paper. In contrast with the
full $B_{i}(\alpha,\beta,u)$, these $\beta\to 0$ singular terms are
\emph{known in full} to the NNLO
~\cite{Gardi:2005yi,Andersen:2005bj,Andersen:2005mj} --- see e.g.
Eq. (3.41) in \cite{Andersen:2005mj} --- and they play an important
r\^ole in Sudakov resummation.

The perturbative expansion in \eq{Ri_borel} corresponding to
\eq{B_i_small_beta} contains log--enhanced terms of the form
$c_i^{(n)}\sim \,\ln^k(\beta)/\beta$, with $k\leq n$. Going beyond
the large--$\beta_0$ limit one finds higher powers of the logarithms
owing to multiple gluon emission: at the $n$-th order one obtains
$\ln^k(\beta)/\beta$ where $k$ goes up to $2n-1$, see \eq{r_mom_def}
and \eq{Sud} below.

\eq{B_i_small_beta} represents the small--$\beta$ limit of all
structure functions except for $i=4$ where the LO vanishes, so the
bremsstrahlung contribution is entirely integrable, $c_4^{(n)}\sim
\,\ln^k(\beta)$, with $k\leq n$. In the latter case we find the
following leading small--$\beta$ behavior:
\begin{align}
\begin{split}
 B_{4}^{\SDG}(\alpha,\beta,u)\simeq\frac{2(\alpha\beta)^{-u}}{\alpha}
\frac{ {\rm e}^{\frac53 u}}{u}\!
\left[
 4(1-u)\left(\frac{\beta}{\alpha}\right)^{-u}\!-\frac{\sin \pi u}{\pi u}\left(\frac{4}{(1-u)^2}
 -\frac{2+\alpha}{1-u}
 +\frac{2+\alpha}{1-u/2}
\right)
 \right].
 \end{split}
\end{align}

The decay width being infrared and collinear safe, the infrared
singularities in \eq{B_i_small_beta} become integrable near the
Sudakov limit once virtual corrections are included. It is therefore
convenient to define an integration prescription absorbing
singularities from virtual corrections into the real emission part.
Defining $r=\beta/\alpha$, we use the plus prescription as follows:
\begin{equation}
\label{plus}
\int_0^{r_0} dr F(r) \left[\frac{1}{r^{1+u}}\right]_{+}=\frac{F(0)}{u}\Big(1-r_0^{-u}\Big)
\,+\,\int_0^{r_0}dr
\Big(F(r)-F(0)\Big)\frac{1}{r^{1+u}},
\end{equation}
where $F(r)$ is a smooth test function. Upon expansion in $u$, this
definition reproduces the $(...)_*$ distribution adopted in
\cite{DeFazio:1999sv,Hoang:2005pj} and the plus distribution of
\cite{Aquila:2005hq}. Equivalently, we can use:
\begin{equation}
\frac1{r^{1+u}} \longrightarrow \left[ \frac1{r^{1+u}} \right]_+-
\frac{\delta(r)}{u}= \left[ \frac1{r^{1+u}} \right]_+- \frac{\alpha
\,\delta(\beta)}{u}.
\end{equation}
Having defined the
integration prescription, the real--emission coefficients
\hbox{$c_i^{(n)}(\alpha,\beta=r\alpha)$}, at each order $n$ in the
expansion (\ref{Ri_borel}), are divided into two parts: the singular
part of (\ref{B_i_small_beta}) is put under a plus prescription, and
the remaining, regular part (which requires no prescription) is left
unmodified:
\begin{equation}
\label{separation} c_i^{(n)}\longrightarrow
\Big[c_i^{(n)\,\sing}\Big]_{+}\,+\, c_i^{(n)\,\reg}.
\end{equation}
Considering in particular the first two orders
$c_i^{(1,2)}(\alpha,\beta=r\alpha)$ of Eqs. (\ref{c1}) and
(\ref{c2}), which are written explicitly in Appendix
\ref{sec:real_explicit}, the part that is put under the plus
prescription is:
\begin{align}
\begin{split}
\label{ci_12_sing} c_i^{(1)\, \sing}(\alpha,\beta=r\alpha)&=\frac{
V_i^{\rm LO}(\alpha)}{\alpha}
\,\bigg[-\frac{\ln(r)}{r}-\frac{7}{4}\frac{1}{r}\bigg]\\
c_i^{(2)\, \sing}(\alpha,\beta=r\alpha)&=\frac{ V_i^{\rm
LO}(\alpha)}{\alpha}
\,\bigg[\frac{3}{2}\frac{\ln^2(r)}{r}+\left(2\ln(\alpha)+\frac{13}{12}\right)
\frac{\ln(r)}{r}\\&+\left(\frac{7}{2}\ln(\alpha)+\frac{\pi^2}{6}-\frac{85}{24}\right)\frac{1}{r}\bigg].
\end{split}
\end{align}
These expressions are consistent with Eq.~(3.41) in
Ref.~\cite{Andersen:2005mj}, where at the NNLO additional terms, with different
color factors, are included.

The subtraction corresponding to this integration prescription will be
applied when regularizing the virtual corrections, see \eq{virt_sing}
and Sec.~\ref{sec:Virt} below. It should be emphasized that defining
plus distributions with respect to $r$ is just a matter of convention,
and depending of the kinematic variables used other choices may turn
out more convenient. In Sec.~\ref{sec:changing_variables} we shortly
discuss two alternatives, demonstrating the way in which different
infrared--subtraction procedures shuffle terms between the real and
virtual contributions.

As discussed in Refs.~\cite{Gardi:2004ia,Gardi:2005yi,Andersen:2005mj},
\eq{B_i_small_beta} contains infrared and collinear singular
contributions that are associated with \emph{two} subprocesses, which
decouple in the Sudakov limit, the quark distribution (in an on-shell
heavy quark) with momentum ${\cal O}(m_b\beta)$ and the jet with
virtuality ${\cal O}(m_b\sqrt{\alpha\beta})$.  The two terms in the
square brackets in \eq{B_i_small_beta} correspond to these two Sudakov
anomalous dimensions, respectively, which were computed here, again,
in the large--$\beta_0$ limit.

To perform Sudakov resummation, accounting for \emph{multiple} soft and collinear radiation,
we follow~\cite{Andersen:2005mj} and define moments of the structure functions with
respect to $r=\beta/\alpha$, in accordance with the plus prescription (\ref{plus}):
\begin{eqnarray}
\label{r_mom_def} {\widetilde{\mathbb{W}}}_i(\alpha,N)
 &\equiv &
  \int_{0}^{\alpha} {d\beta}
 \left(1-\frac{\beta}{\alpha}\right)^{N-1}
{\mathbb{W}}_i(\alpha,\beta)
 \nonumber \\
&=& H_i(\alpha)\times
{\rm Sud}(m_b\alpha,N)+\Delta R_i^{(N)}(\alpha),
\end{eqnarray}
where $H_i(\alpha)=\left.V_i(\alpha)\right\vert_{{\rm large}\,\,\beta_0}+\cdots$.
We can deduce the structure of the Sudakov exponent from
\eq{B_i_small_beta}:
\begin{eqnarray}
\label{Sud}
 \hspace*{-20pt}{\rm Sud}(m_b\alpha,N)&=&
\exp\Bigg\{\frac{C_F}{\beta_0}\, \int_0^{\infty}\frac{du}{u}\,T(u)\,
 \left(\frac{\Lambda^2}{\alpha^2 m_b^2}\right)^u
\Bigg[B_{\cal S}(u)
\left(\frac{\Gamma(N)\Gamma(-2u)}{\Gamma(N-2u)}+\frac{1}{2u}\right)
\\\nonumber &&\hspace*{170pt}
 -B_{\cal J}(u)\left(\frac{\Gamma(N)\Gamma(-u)}{\Gamma(N-u)}
 +\frac{1}{u}
\right)
 \Bigg]\Bigg\},
\end{eqnarray}
which we wrote as a Borel integral (in the DGE form) with
\begin{eqnarray}
\label{B_JS}
\left.B_{\cal J}(u)\right\vert_{\rm large\,\,\beta_0}
&=&{\rm e}^{\frac53 u}
\frac12\frac{\sin \pi u}{\pi u}\left(\frac{1}{1-u}+\frac{1}{1-u/2}
\right)\\
\left.B_{\cal S}(u)\right\vert_{\rm large\,\,\beta_0}&=&
{\rm e}^{\frac53 u} (1-u).
\end{eqnarray}
In the exponent in \eq{Sud} we \emph{added}, under the Borel integral
\[
\frac{C_F}{\beta_0}\int_0^{\infty}du \,T(u)\,
 \left(\frac{\Lambda^2}{m_b^2}\right)^u \times \Bigg[\,\,\ldots\,\, \Bigg]
\]
the singularities that are required for writing the $r\to 0$
non-integrable terms of \eq{B_i_small_beta} as a plus distribution
(\eq{plus}):
\begin{eqnarray}
\label{virt_sing}
\hspace*{-3pt}\left.B[V_0](\alpha,u)\right\vert_{\rm sing.}&\equiv&
\frac{\alpha^{-2u}}{2u^2}\bigg[B_{\cal S}(u)-2B_{\cal J}(u)
\bigg]
\\ \nonumber
&=&
\frac{\alpha^{-2u}\,{\rm e}^{\frac53 u}}{2u^2} \,
\Bigg[(1-u)  -\frac{\sin \pi u}{\pi u}\left(\frac{1}{1-u}+\frac{1}{1-u/2}
\right)\Bigg]
 \\  \nonumber
&=& -\frac{1}{2u^{2}}\,+\,\left(\ln(\alpha)-\frac{25}{12}\right)\frac{1}{u}-
\left(\ln^2(\alpha)-\frac{25}{6}\ln(\alpha)-\frac{1}{6}\pi^2+\frac{245}{72}\right)
+{\cal O}(u),
\end{eqnarray}
 making the moments in Eqs.
(\ref{r_mom_def}) and (\ref{Sud}) above finite. These terms will be
\emph{subtracted}
from the virtual corrections in \eq{BV_reg} below. As shown explicitly in Sec.~\ref{sec:Virt}
this subtraction exactly cancels the infrared singularities of the virtual corrections.

\section{Virtual corrections\label{sec:Virt}}

The virtual contribution to the structure functions\footnote{Our
Lorentz decomposition is similar to that of Appendix B
in~Ref.~\cite{Aquila:2005hq}; it differs from that of
Ref.~\cite{DeFazio:1999sv}.} can be decomposed as
in~\eq{StructureFunctions}:
\begin{align}
\begin{split}
 {V}^{\mu\nu}(p, q) &=-  {V}_1(\alpha) g^{\mu\nu}
+ {V}_2(\alpha)v^{\mu}v^{\nu} +
i {V}_3(\alpha)\epsilon^{\mu\nu\rho\sigma}v_{\rho}\hat{q}_{\sigma}
+ {V}_4(\alpha)\hat{q}_{\mu} \hat{q}_{\nu}
+ {V}_5(\alpha) (v^{\mu}\hat{q}^{\nu} +
v^{\nu}\hat{q}^{\mu}).
\end{split}
\end{align}
For each structure function one has an expansion in the coupling, see
\eq{Borel_virt} below; according to (\ref{LO}), at leading order (Born level)
we have:
\begin{align}
\label{V_LO}
\begin{split}
 {V}^{\mu\nu}_{\rm LO}(p, q) &=-  \alpha g^{\mu\nu}
+ 4v^{\mu}v^{\nu} +
i 2\epsilon^{\mu\nu\rho\sigma}v_{\rho}\hat{q}_{\sigma}
 -2 (v^{\mu}\hat{q}^{\nu} +v^{\nu}\hat{q}^{\mu}).
\end{split}
\end{align}

Let us now compute the virtual corrections in the large--$\beta_0$
limit. To this end we modify the gluon propagator according to
\begin{equation}
\label{prop}
\frac{g_{\mu\nu}}{-k^2}\,\longrightarrow\,\frac{g_{\mu\nu}}{(-k^2)^{1+u}}.
\end{equation}
With this modification\footnote{See \cite{Beneke:1998ui} or Sec. 2.2
in \cite{Gardi:2006rp}.}, the momentum integration should yield
directly the Borel function $B[V_i](\alpha,u)$ in
\begin{eqnarray}
\label{Borel_virt} V_i^{\SDG}(\alpha)&\bumpeq& V_i^{\LO}(\alpha)\,+\,
\frac{C_F}{\beta_0}
\int_0^{\infty}du\,T(u)\,\left(\frac{\Lambda^2}{m_b^2}\right)^u\,
B[{V}_i](\alpha,u)
\end{eqnarray}
However, in contrast with the
real--emission result of \eq{Ri_borel} that is regular for \hbox{$u\to 0$},
the Borel integral of the virtual diagrams is obstructed by a double
pole of $B[{V}_i](\alpha,u)$ at $u\to 0$, which corresponds to the usual double--logarithmic
infrared singularity. Therefore, after computing the momentum
integral we shall perform infrared subtraction using the singular
terms (\ref{virt_sing}). This would finally yield a meaningful Borel
representation for the virtual contribution, \eq{Vi_reg_borel} below.

We define
$z\equiv q^2/m_b^2$,
and in the following, since $\beta=0$, we have $z=1-\alpha$. The result of
the virtual diagrams, where the gluon propagator is modified according to
(\ref{prop}), takes the form\footnote{This result agrees with
Eqs. of (B.11) and (B.12)
in~Ref.~\cite{Aquila:2005hq} (with $m_c=0$). There a gluon mass
is introduced instead of a Borel parameter.}:
\begin{align}
\label{BV}
\begin{split}
 B[{V}^{\mu\nu}](p, q) &= B[V_0](\alpha,u)\,\times\, {V}^{\mu\nu}_{\rm LO}(p, q)\\&
 -  B[{V}_1](\alpha,u)\, g^{\mu\nu}\\&
+ B[{V}_2](\alpha,u)\,v^{\mu}v^{\nu} \\&+
i B[{V}_3](\alpha,u)\,\epsilon^{\mu\nu\rho\sigma}v_{\rho}\hat{q}_{\sigma}
\\&+ B[{V}_4](\alpha,u)\,\hat{q}_{\mu} \hat{q}_{\nu}
\\&+ B[{V}_5](\alpha,u)\, (v^{\mu}\hat{q}^{\nu} +
v^{\nu}\hat{q}^{\mu})
\end{split}
\end{align}
where
\begin{eqnarray}
\label{V_I}
B[V_0](\alpha,u)&=& {\rm e}^{\frac53 u}\left[
\frac12 D(u)+\frac12 C(z,u) +(1-z) \Big(I_x(z,u) +I_y(z,u)-I_1(z,u)\Big)\right];\nonumber\\
B[V_1](\alpha,u)&=&-{\rm e}^{\frac53 u}(1-z)\Big(K(z,u)-I_x(z,u)\Big); \nonumber\\
B[V_2](\alpha,u)&=&4\,{\rm e}^{\frac53 u}\,z\, I_{xy}(z,u); \nonumber\\
B[V_3](\alpha,u)&=&-2\,{\rm e}^{\frac53 u}\,\Big(K(z,u)-I_x(z,u)\Big);\nonumber\\
B[V_4](\alpha,u)&=&-4\,{\rm e}^{\frac53 u}\,\Big(I_x(z,u) -I_{xy}(z,u)\Big); \nonumber\\
B[V_5](\alpha,u)&=&-2\,{\rm e}^{\frac53 u}\,\Big[(1+z)I_{xy}(z,u)-I_x(z,u)\Big].
\end{eqnarray}

In \eq{V_I} the term $D(u)$
\begin{eqnarray}
D(u)&=&\int^1_0 dx (1-x)^{1+u}x^{-2u} \left( 2\frac{1+x}{x} -\frac1u \right)=
 -3\frac{1-u}{u}\,\frac{\Gamma(2+u)\Gamma(1-2u)}{\Gamma(3-u)}
\nonumber\\
&=&-\frac3{2u}-\frac94-u\left(\frac98 +\frac{\pi^2}{2}\right)+O(u^2)
\end{eqnarray}
is the result of the $b$-quark self--energy diagram\footnote{Our result for the self--energy diagram
agrees with that of Ref.~\cite{Beneke:1994sw}.}, while the $u$-quark
self--energy diagram vanishes in the Borel regularization, as the
momentum integral has no scale. All the remaining terms in (\ref{V_I})
arise from the vertex correction diagram. Let us recall that the Borel
parameter regularizes both ultraviolet and infrared
logarithmic singularities, just as in dimensional regularization.
Thus, $u\to 0$ singularities in individual diagrams arise from both the
ultraviolet and the infrared and no distinction is made between them.
However, in the present context we know in advance that the ultraviolet
divergencies cancel out in the sum of all diagrams ---
the current is conserved --- and therefore the remaining
$u\to 0$ singularities in the sum of diagrams are immediately identified as
infrared ones. We will address these singularities below.

Let us now briefly describe the calculation of the vertex diagram
and define the integrals entering~\eq{V_I}.
Upon combining the propagators using  Feynman parametrization,
where the $b$--quark propagator is associated with the Feynman parameter $x$ and
the $u$--quark propagator with $y$ (so the gluon with $1-x-y$),
one identifies the scale
 \begin{eqnarray}
M^2\equiv m_b^2 \,x\Big(y (1-z)+x \Big),
\end{eqnarray}
where $z\equiv q^2/m_b^2$ as above.
Performing next the loop--momentum integral in four dimensions one
obtains integrals of the following form over the Feynman parameters:
\begin{equation}
I_{a,b,c}(z,u)\equiv \int_0^1 dx \int_0^{1-x} dy \frac{x^a \,y^b \,(1-x-y)^u}
{\Big[x\Big(y(1-z)+x\Big)\Big]^{c+u}},
\end{equation}
where $a$, $b$ and $c$ are non-negative integers.
This integral is computed as follows. One first changes variables into
from $y$ into $w=1-x-y$ and then from $x$ into $t$ where $x=(1-w)(1-t)$. The integration over
both $w$ and $t$ extends over the interval $[0,1]$.
In these variables the integrand factorizes and, assuming that the parameters $a$, $b$ and $c$
are such that both integrals exist (which is always the case for the required integrals) the result is:
\begin{eqnarray}
I_{a,b,c}(z,u)&=& \frac{\Gamma(a-c-u+1) \Gamma(b+1)\Gamma(1+u) \Gamma(2+a-2 c-2 u+b)}
{\Gamma(2+a-c-u+b)\Gamma(3+a+b-2 c-u)} \times\nonumber
\\&&\, _2F_1([c+u, b+1],[2+a-c-u+b],z).
\end{eqnarray}
At the next step, known hypergeometric identities (see e.g.~\cite{Kalmykov:2006pu})
are used to bring the result into one that
is convenient to expand at small $u$. In all cases it is possible to write the result in terms
of a single hypergeometric function $_2F_1([1, 1+u],[2-u],z)$.
This function has the following expansion (type E in Ref.~\cite{Kalmykov:2006pu}):
\begin{align}
\begin{split}
_2F_1([1, 1+u],[2-u],z)&\simeq
\frac{1-u}{z}\, \Bigg\{-\ln(1-z)-u \,\Big(-\ln^2(1-z)-{\rm Li}_2(z)\Big)
\\&\hspace{-40pt}+u^2\Big(-2 S_{1,2}(z)-2 \ln(1-z)
{\rm Li}_2(z)+{\rm Li}_3(z)-\frac{2}{3} \ln^3(1-z)\Big)\\&\hspace{-40pt}-u^3\,
\Big(2 S_{2,2}(z)+2 \ln(1-z) {\rm Li}_3(z)
-4 \ln(1-z) S_{1,2}(z)\\&\hspace{-40pt}-2 \ln^2(1-z) {\rm Li}_2(z)-\frac{1}{3}
\ln^4(1-z)-4 S_{1,3}(z)-{\rm Li}_4(z)\Big)
+{\cal O}(u^4)
\Bigg\}.
\end{split}
\end{align}

Following Ref.~\cite{Aquila:2005hq} (see Eq. (B.3) there) we separate the
numerator of the vertex diagram into $N_1^{\mu\nu}$, which is composed of
terms having powers of the loop momentum in the numerator
({\it cf.} Eq. (B.6) in Ref.~\cite{Aquila:2005hq}) and
other terms, $N_2^{\mu\nu}$ ({\it cf.} Eq. (B.9)
in Ref.~\cite{Aquila:2005hq}).
In \eq{V_I} above, the $N_1^{\mu\nu}$ gives rise to $C(z,u)$ in
$B[V_0](\alpha,u)$:
\begin{eqnarray}
C(z,u)= \frac{2}{u} K(z,u),
\end{eqnarray}
where
\begin{align}
\begin{split}
K(z,u)&=
\int^1_0 dx \int_0^{1-x} dy \frac{m_b^{2u}(1-x-y)^u}{(M^2)^u}
\\
&= \frac{\Gamma(1-2u)\Gamma(1+u)}{\Gamma(3-u)}
\left[1-\frac{u(1-z)}{1-u} \, _2F_1([1,1+u],[2-u],z)\right]
\\
&= \frac12 +u\left(\frac34 +\frac{1-z}{2z}\ln(1-z)\right)+O(u^2)\,,
\end{split}
\end{align}
while the
$N_2^{\mu\nu}$ is the source of all the other terms, where the following
additional integrals show up:
\begin{align}
\begin{split}
I_1(z,u)&= \int^1_0 dx \int_0^{1-x} dy \frac{m_b^{2u}(1-x-y)^u}{(M^2)^{1+u}}\\
        &=
        \frac{\Gamma(1+u)\Gamma(1-2u)}{2\,u^2\Gamma(1-u)\,(1-z)}
        \left[1+\frac{2u\,z}{1-u}\, _2F_1([1, 1+u],[2-u],z)\right]
        \\
        &=\frac1{1-z}\Bigg\{
\frac{1}{2\,u^2}\,-\,\frac{\ln(1-z)}{u}
\,+\,\frac{\pi^2}{6}+\ln^2(1-z)+{\rm Li}_2(z)
-\frac{u}{3}\Big(\ln(1-z)\pi^2
\\&-3\zeta_3+6 S_{1,2}(z)+6\ln(1-z){\rm Li}_2(z)-3{\rm Li}_3(z)
+2\ln^3(1-z)\Big)
+{\cal O}(u^2)\Bigg\},
\end{split}
\end{align}
\begin{align}
\begin{split}
I_x(z,u)&= \int^1_0 dx \int_0^{1-x} dy \frac{m_b^{2u}(1-x-y)^u\, x}{(M^2)^{1+u}}
\\ &=
\frac{\Gamma(1-2u)\Gamma(1+u)}{(1-u)\Gamma(2-u)}
\, _2F_1([1,1+u],[2-u],z)
\\   &= -\frac{\ln (1-z)}{z}+
\frac{1}{z}\Big(\ln^2(1-z)+{\rm Li}_2(z)-\ln(1-z)\Big)\,u
+{\cal O}(u^2),
\end{split}
\end{align}
\begin{align}
\begin{split}
I_y(z,u)&= \int^1_0 dx \int_0^{1-x} dy \frac{m_b^{2u}(1-x-y)^u\, y}{(M^2)^{1+u}}\\
        &=-\frac{\Gamma(1+u)\Gamma(1-2u)}{u\,\Gamma(2-u)(1-z)}
            \left[1+\frac{u(1+z)}{1-u}\, _2F_1([1, 1+u],[2-u],z)\right]\\
        &=
            \frac{1}{1-z}\Bigg\{
            -\frac{1}{u}-1+\left(1+\frac{1}{z}\right)\ln(1-z)
            \\
        &+\left[\left(1+\frac{1}{z}\right)\left(-\ln^2(1-z)+
            \ln(1-z)-{\rm Li}_2(z)\right)-
            \frac{1}{3}\pi^2-1\right]\,u\, +{\cal O}(u^2)\Bigg\},
\end{split}
\end{align}
and
\begin{align}
\begin{split}
I_{xy}(z,u)&= \int^1_0 dx \int_0^{1-x} dy \frac{m_b^{2u}(1-x-y)^u\, xy}{(M^2)^{1+u}}\\
      &=
      -\frac{\Gamma(1+u)\Gamma(1-2u)}{z\Gamma(3-u)}
      \left[1-\left(1+z-\frac{z}{1-u}\right)\, _2F_1([1, 1+u],[2-u],z)\right]\\
      &=
      \frac{1}{2z^2}\Bigg\{-\ln(1-z)-z+
      \bigg[\ln^2(1-z)+\left(z-\frac{1}{2}\right)\ln(1-z)\\
      &-\frac{3}{2}z+{\rm Li}_2(z)\bigg]\,u
      +{\cal O}(u^2) \Bigg\}.
\end{split}
\end{align}

\eq{BV} is written such that all the $u\to 0$ infrared singularities
are in the first term.
This term is proportional to the LO result, as must be the case.
As mentioned above, in the absence of such singularities
one would interpret the expansion
of the virtual corrections, starting at ${\cal O}(\alpha_s)$,
according to \eq{Borel_virt} above. Obviously, since there is a
double pole at $u=0$, the $u$-integral (\ref{Borel_virt}) is ill--defined.
One should first perform subtraction of the singularities.

Using the explicit results for the integrals given above, it is straightforward
to check that the $u\to 0$ singularities of $B[V_0](\alpha,u)$ in \eq{V_I}
do indeed coincide with \eq{virt_sing}, that was determined in Sec.~\ref{sec:Sudakov}
by defining the plus distribution for the real--emission terms.
After subtracting from $B[V_0](\alpha,u)$ in \eq{V_I} the terms
in \eq{virt_sing} one has
\begin{align}
\label{BV_reg}
\begin{split}
\left. B[{V}^{\mu\nu}](p, q)\right\vert_{\rm reg.} &= \bigg[B[V_0](\alpha,u)-\,
\left.B[V_0](\alpha,u)\right\vert_{\rm sing.}\bigg]\,\times\, {V}^{\mu\nu}_{\rm LO}(p, q)\\&
 -  B[{V}_1](\alpha,u)\, g^{\mu\nu}\\&
+ B[{V}_2](\alpha,u)\,v^{\mu}v^{\nu} \\&+
i B[{V}_3](\alpha,u)\,\epsilon^{\mu\nu\rho\sigma}v_{\rho}\hat{q}_{\sigma}
\\&+ B[{V}_4](\alpha,u)\,\hat{q}_{\mu} \hat{q}_{\nu}
\\&+ B[{V}_5](\alpha,u)\, (v^{\mu}\hat{q}^{\nu} +
v^{\nu}\hat{q}^{\mu}).
\end{split}
\end{align}
Next, let us split the regularized terms proportional to
${V}^{\mu\nu}_{\rm LO}$ using \eq{V_LO}
and absorb them into the five different structure functions; we define:
\begin{align}
\label{BV_reg_}
\begin{split}
\left. B[{V}^{\mu\nu}](p, q)\right\vert_{\rm reg.} &=
 -  \left. B[{V}_1](\alpha,u)\right\vert_{\rm reg.}\, g^{\mu\nu}\\&
+ \left.B[{V}_2](\alpha,u)\right\vert_{\rm reg.}\,v^{\mu}v^{\nu} \\&+
i \left.B[{V}_3](\alpha,u)\right\vert_{\rm reg.}\,\epsilon^{\mu\nu\rho\sigma}v_{\rho}\hat{q}_{\sigma}
\\&+ \left.B[{V}_4](\alpha,u)\right\vert_{\rm reg.}\,\hat{q}_{\mu} \hat{q}_{\nu}
\\&+ \left. B[{V}_5](\alpha,u)\right\vert_{\rm reg.}\, (v^{\mu}\hat{q}^{\nu} +
v^{\nu}\hat{q}^{\mu}).
\end{split}
\end{align}
Finally, using \eq{V_I} with the explicit results for the integrals we get:
\begin{align}
\label{BV_reg_expl}
\begin{split}
\left.B[{V}_1](\alpha,u)\right\vert_{\rm reg.} &=\alpha\,{\rm e}^{\frac53 u}
\Bigg\{ \frac{ (2 \alpha+u-2)}{u-1}\,
 \mathbb{G}(u) \, \mathbb{F}(u,\alpha)
+ \left(\frac{1}{u}+1+\frac{1}{2}u-\frac{3}{2}u^2\right) \,\mathbb{G}(u)\\&
-\frac{\alpha^{-2u}}{2u^2} \,
\bigg[1-u  -\frac{\sin \pi u}{\pi u}\left(\frac{1}{1-u}+\frac{1}{1-u/2}
\right)\bigg]\Bigg\},
\\
\left.B[{V}_2](\alpha,u)\right\vert_{\rm reg.}&=
\frac{4}{\alpha}\,\left.B[{V}_1](\alpha,u)\right\vert_{\rm reg.}+
\frac{4\, {\rm e}^{\frac53 u}\, u \,(1+u)}{1-u}\,  \mathbb{G}(u)
\, \mathbb{F}(u,\alpha)
\\
\left.B[{V}_3](\alpha,u)\right\vert_{\rm reg.}&=
\frac{2}{\alpha}\,\left.B[{V}_1](\alpha,u)\right\vert_{\rm reg.}
\\
\left.B[{V}_4](\alpha,u)\right\vert_{\rm reg.}&=
\frac{4\,u}{1-\alpha}\, {\rm e}^{\frac53 u}\,\mathbb{G}(u)\,\bigg[
\frac{ (2 \alpha-u-1)}{u-1} \,
 \mathbb{F}(u,\alpha)
 +1
\bigg],
\\
\left.B[{V}_5](\alpha,u)\right\vert_{\rm reg.}&=
-\frac{2}{\alpha}\,\left.B[{V}_1](\alpha,u)\right\vert_{\rm reg.}+
\frac{2\,u\,{\rm e}^{\frac53 u}}{1-\alpha}\,\mathbb{G}(u)\,
\left[\frac{ 2 (u+1)-\alpha\, (u+3)}{u-1}\,
 \mathbb{F}(u,\alpha)-1\right]\,,
\end{split}
\end{align}
where ${\mathbb{G}(u)} \equiv -\Gamma(1-2 u)\Gamma(u)/\Gamma(3-u)$
and ${\mathbb{F}(u,\alpha)} \equiv \, _2F_1([1, \,1 + u], \,[2 - u], \,1 - \alpha )$.

The infrared--subtracted virtual terms can be expanded order by
order in $u$ to get the perturbative corrections in the
large--$\beta_0$ limit. Let us write, in analogy with the real--emission
result (\ref{Ri_borel}), a Borel representation of $V_i(\alpha)$ in
(\ref{W_RV}):
\begin{eqnarray}
\label{Vi_reg_borel} V_i^{\SDG}(\alpha)&=& V_i^{\LO}(\alpha)\,+\,
\frac{C_F}{\beta_0}
\int_0^{\infty}du\,T(u)\,\left(\frac{\Lambda^2}{m_b^2}\right)^u\,
\left.B[{V}_i](\alpha,u)\right\vert_{\rm reg.}\\ &=&
V_i^{\LO}(\alpha)\,+\,C_F\,\bigg[v_i^{(1)}(\alpha)\,\frac{\alpha_s(m_b)}{\pi}+
v_i^{(2)}(\alpha)\,\beta_0\left(\frac{\alpha_s(m_b)}{\pi}\right)^2+\cdots\bigg].
\nonumber
\end{eqnarray}
The coefficients $v_i^{(1,2)}(\alpha)$ for the five structure
functions, $i=1$ to $5$ are listed in
Appendix~\ref{sec:Virtual_NLO_NNLO}.

\section{The triple--differential width at NLLO in the large--$\beta_0$ limit
\label{sec:put_together}}

In the previous sections we computed separately real and virtual
corrections to the five different structure functions in the
decomposition of the hadronic tensor. For massless leptons, only three
of those enter into the expression for the spectrum
(\ref{triple_diff_light_cone_alpha_beta_x}) through (\ref{WL}). Let us
now combine the result into an expression for the triple--differential
width. We present explicit expressions up to NNLO which are valid in
the on-shell quark mass scheme.  Using the results of the previous sections the
generalization to higher orders is straightforward.

Let us write the perturbative  expansion of the triple--differential width
in the large--$\beta_0$ limit  as follows:
\begin{align}
\label{full_result_expanded}
\begin{split}
\frac{1}{\Gamma_0}\,\frac{d^3\Gamma}{d\alpha\,d\beta\, d x}&=
\omega_0(\alpha, x)\,\delta(\beta)
\\& +C_F \Bigg[\frac{\alpha_s(m_b)}{\pi}\, K_1(\alpha,\beta, x)
+ \,\beta_0\,
\left(\frac{\alpha_s(m_b)}{\pi}\right)^2\,K_2(\alpha,\beta, x)
+\cdots\bigg],
\end{split}
\end{align}
where $\omega_0(\alpha, x)$ is defined in (\ref{LO_result})
and the NLO and NNLO coefficients $K_{n}(\alpha,\beta, x)$ for $n=1$ and $2$,
respectively, will be detailed below.
At each order real and virtual contributions to each structure function add up
according to (\ref{W_RV}).  The coefficients of the triple differential width can
therefore be written as follows:
\begin{align}
\label{NiLO_result}
\begin{split}
K_n(\alpha,\beta, x)&=
\omega_n(\alpha,x)\,\delta(\beta)+
\Big\{K_n^{\sing}(\alpha,\beta, x)\Big\}_{+}\,+\,
K_n^{\reg}(\alpha,\beta, x),
\end{split}
\end{align}
where, as usual $\beta=\alpha\, r$ and the plus prescription is as defined with respect to $r$
according to Eq.~(\ref{plus}).

To obtain the virtual coefficients $\omega_i(\alpha,x)$ at each order $n$
one substitutes the regularized virtual coefficients of the structure functions
(\ref{Vi_reg_borel}),
which are given explicitly in Appendix~\ref{sec:Virtual_NLO_NNLO} for $n=1,\,2$, into
Eq.~(\ref{WL}) and uses the result in~(\ref{triple_diff_light_cone_alpha_beta_x}).
The virtual coefficient at NLO  ($n=1$) is
\begin{align}
\label{NLO_result_V}
\begin{split}
\omega_1(\alpha,x)\,=\, &
 - \omega_0(\alpha,x)\bigg(\mathrm{
Li}_2(1 - \alpha ) + \frac54 + \frac{\pi ^{2}}{3}\bigg) +6\,(\alpha
- 1 + x)\, (2\,\alpha  - 5 + 2\,x)\,\mathrm{ln}\,(\alpha)
\end{split}
\end{align}
and the NNLO result ($n=2$) is:
\begin{align}
\label{NNLO_result_V}
\begin{split}
\omega_2(\alpha,x)
&\,=\, 6 \,(\alpha  - 1 + x)\left[ \mathrm{Li}_2(1 - \alpha )
+ \mathrm{ln}^{2}(\alpha) - {\displaystyle \frac {25}{6}} \,
\mathrm{ln}(\alpha)\right]
\\& \hspace{-10mm}
-\omega_0(\alpha,x) \,    
\bigg[\text{Li}_3(1-\alpha)+2 \text{Li}_3(\alpha)+
\frac{19}{6}\text{Li}_2(1-\alpha)
\\& \hspace{-10mm}
+( \ln (1-\alpha)-1) \ln ^2(\alpha)
+\left(\frac{4-7 \alpha}{6(1- \alpha)}- \pi ^2\right) \ln (\alpha)
- \zeta_3+
\frac{79 \pi^2}{72}+\frac{71}{24}
 \bigg].
\end{split}
\end{align}

Similarly, by substituting the real--emission results for the structure functions into (\ref{WL}),
one obtains the corresponding real--emission coefficients for the triple differential width.
The singular part, $K_n^{\sing}(\alpha,\beta, x)$, which
enters (\ref{NiLO_result}) under a plus prescription (\ref{plus}), is obtained by
\begin{align}
\label{NnLO_result_sing}
\begin{split}
K_n^{\sing}(\alpha,\alpha\,r, x)
\,=\,& 6\, (1-\alpha)
(1-\alpha\, r)\,c_1^{(n)\,\sing}(\alpha,\alpha\, r)
 \\& \hspace*{-40pt}
-3\,\bigg(x^2\,- \,x \, (2-\alpha-\alpha\, r)+(1-\alpha) (1-\alpha\, r)
 \bigg)\,
c_2^{(n)\,\sing}(\alpha,\alpha\, r)\\& \hspace*{-40pt}
+\,
 6\,(1-\alpha) (1-\alpha\, r) \left(x-1+\frac12 (\alpha+\alpha\, r)\right)\,
c_3^{(n)\,\sing}(\alpha,\alpha\, r)\,.
\end{split}
\end{align}
Since at any order $n$ the coefficients $c_i^{(n)\,\sing}(\alpha,\alpha\, r)$
are proportional to the corresponding
LO result $V_i^{\rm LO}(\alpha)$ for $i=1$ to $3$, one obtains
the singular part $K_n^{\sing}(\alpha,\beta, x)$ as the $r\to 0$ singular terms
corresponding to the expansion of
(\ref{B_i_small_beta}) in powers of $u$, which depend only on $r$ and $\alpha$, times
the following prefactor,
\begin{equation}
\label{eq:Omega}
\Omega(\alpha,\,r,\,x)\equiv {\omega_0(\alpha,x)}
+6\Big(2\alpha^2-7\alpha-4x+5+2x\alpha\Big)\, \alpha r
-6\alpha^2 (1-\alpha)\,\,r^2,
\end{equation}
that depends also on the lepton energy fraction $x$.
In particular, using \eq{ci_12_sing}, the NLO result is
\begin{align}
\label{NLO_result_sing}
\begin{split}
K_1^{\sing}(\alpha,\,\alpha\,r,\, x)
\,=\,& \frac{\Omega(\alpha,\,r,\,x)}{\alpha}\,\left[-\frac{\ln(r)}{r}-\frac{7}{4}\,\frac{1}{r}\right]\,,
\end{split}
\end{align}
and the NNLO one is:
\begin{align}
\label{NNLO_result_sing}
\begin{split}
K_2^{\sing}(\alpha,\alpha\,r, x)
&\!=\!
\frac{\Omega(\alpha,\,r,\,x)}{\alpha}\,
\bigg[\frac{3}{2}\frac{\ln^2(r)}{r}+\left(2\ln(\alpha)+\frac{13}{12}\right)
\frac{\ln(r)}{r}+\left(\frac{7}{2}\ln(\alpha)+\frac{\pi^2}{6}-\frac{85}{24}\right)
\frac{1}{r}\bigg].
\end{split}
\end{align}
As expected, the ${\cal O}(r^0)$ term in $\Omega(\alpha,\,r,\,x)$ coincides with
the Born--level result $\omega_0(\alpha,x)$.
However, owing to the contraction with the leptonic tensor in (\ref{NnLO_result_sing}),
$\Omega(\alpha,\,r,\,x)$ also contains some ${\cal O}(r^1)$ and ${\cal O}(r^2)$ terms
that  generate  integrable ${\cal O}(r^0)$ and ${\cal O}(r^1)$ terms
in $K_n^{\sing}(\alpha,\alpha\,r, x)$.
These terms can be freely taken out of the $\left\{...\right\}_+$ brackets
in~(\ref{NiLO_result}), as they do not vary by applying the plus prescription
(\ref{plus}).

Finally, the regular terms $K_n^{\reg}(\alpha,\beta, x)$ are obtained by
\begin{align}
\label{NnLO_result_reg}
\begin{split}
K_n^{\reg}(\alpha,\alpha\,r, x)
\,=\,&   6\,(1-\alpha)
(1-\alpha\, r)\,\Big[c_1^{(n)}(\alpha,\alpha\, r)-c_1^{(n)\,\sing}(\alpha,\alpha\, r)\Big]
 \\& \hspace*{-60pt}
-3\,\bigg(x^2\,- \,x \, (2-\alpha-\alpha\, r)+(1-\alpha) (1-\alpha\, r)
 \bigg)\,
 \Big[c_2^{(n)}(\alpha,\alpha\, r)-c_2^{(n)\,\sing}(\alpha,\alpha\, r)\Big]\\&
\hspace*{-60pt}+\,
 6\,(1-\alpha) (1-\alpha\, r) \left(x-1+\frac12 (\alpha+\alpha\, r)\right)\,
 \Big[c_3^{(n)}(\alpha,\alpha\, r)-c_3^{(n)\,\sing}(\alpha,\alpha\, r)\Big],
\end{split}
\end{align}
where, for $n=1$ and $2$, the explicit expressions for
$c_i^{(n)}(\alpha,\beta)$ and its singular part
are given in Appendix \ref{sec:real_explicit} and in
\eq{ci_12_sing}, respectively. Using these expressions we find that
the regular term at NLO is given by:
\begin{align}
\label{NLO_result_reg}
\begin{split}
K_1^{\reg}(\alpha,\alpha\,r, x)
\,&= \frac{6\,\mathrm{ln}(r)}{\alpha \,(1 - r)^{4}}\,
{\mathbb Q}_1(\alpha,\, r,\, x)
   \\ & \hspace*{-40pt}
\mbox{} - {\displaystyle \frac{3(  1 - \alpha )\,(  1 - \alpha\,r)}
{2\,(  1 - r)} {\,{\mathbb Q}_2(\alpha,\, r,\, x)}}   
-\frac{3\,(1-x -\alpha )\,(1-x - \alpha \,r) }{(  1 - r)
^{3}\,\alpha }\,{\mathbb Q}_3(\alpha,\, r,\, x)   \\ & \hspace*{-40pt}
\mbox{} - {\displaystyle \frac {3\,(  1 - \alpha )\,(  1 - \alpha
\,r)\,(2\,x - 2 + \alpha  + \alpha \,r)\,}{2\,\alpha \,(  1 - r)}\,(4\,\alpha ^{2}\,r - 10
\,\alpha \,r + 7\,r - 10\,\alpha  + 9)},
\end{split}
\end{align}
and at NNLO by
\begin{align}
\label{NNLO_result_reg}
\begin{split}
K_2^{\reg}(\alpha,\alpha\,r, x) \,&= \,-2 \,
K_1^{\reg}(\alpha,\alpha\,r, x) \, \ln (\alpha)
 - 3 \,{\displaystyle \frac {( 1 + r - \alpha \,r)
\,\mathrm{ln}(  { 1 + r -\alpha \,r} )}
{\alpha \,r\,(  1 - \alpha
 )\,(  1 - \alpha \,r)\,(  1 - r)^{3}} \,{\mathbb{P}_{1}(\alpha,\, r,\,x)}} \\& \hspace*{-40pt}
+ {\displaystyle
\frac {1}{2}} \,{\displaystyle \frac {\mathrm{ln}(r)\,{
\mathbb{P}_{2}(\alpha,\, r,\,x)}}{(1 - \alpha \,r)\,\alpha \,(1 - r)^{4}}}
- {\displaystyle \frac {9\,{\mathbb{Q}_{1}(\alpha,\, r,\,x)}}{
\alpha \,(1 - r)^{4}}} \, \mathrm{ln}^{2}(r)
+ \left(\frac{6\,{\mathbb{Q}_{1}(\alpha,\, r,\,x)}}{\alpha \,(1 - r)^{4}} -
\frac{\Omega(\alpha,\,r,\,x)}
{\alpha \,r}\right)\,\times \\ & \hspace*{-40pt}
\times\bigg[ {\mathrm{ln}(
{ 1 + r - \alpha \,r} )\,\mathrm{ln}(r)\,
}  +  {\rm Li}_2\left({\displaystyle \frac {r\,(  1 - \alpha
 )}{ 1 + r - \alpha \,r}} \right) - {\rm Li}_2
 \left({\displaystyle \frac {  1 - \alpha \,r}{ 1 + r - \alpha \,r}}
\right)\bigg]
\\ & \hspace*{-40pt}
+ {\displaystyle \frac {1}{4}} \,
{\displaystyle \frac {{\mathbb{P}_{3}(\alpha,\, r,\,x)}}{(1 - r)}}
 + {\displaystyle \frac {1}{2}} \,{\displaystyle \frac {(
1 - x - \alpha \,r)\,(1 - x - \alpha )}{(1
 - r)^{3}\,\alpha }}\,{\mathbb{P}_{4}(\alpha,\, r,\,x)}  \\ & \hspace*{-40pt}
\mbox{} + {\displaystyle \frac {1}{4}} \,{\displaystyle \frac {(
2\,x - 2 + \alpha  + \alpha \,r)\,(1 - \alpha \,r)\,(1 - \alpha )
\,}{(1 - r)\,\alpha }}\,{\mathbb{P}_{5}(\alpha,\, r,\,x)}-
\frac{\pi^2}{6} \frac{\Omega(\alpha,\,r,\,x)}{\alpha \,r}
\,,
\end{split}
\end{align}
where $\Omega(\alpha,\,r,\,x)$ is given in \eq{eq:Omega} and the polynomials
$\mathbb{Q}_{j}(\alpha,\, r,\,x)$
for $j=1$ to $3$ and $\mathbb{P}_{j}(\alpha,\, r,\,x)$
for $j=1$ to $5$  are listed in Appendix~\ref{sec:Polylomials_NNLO}.

\section{Changing variables: alternative subtraction procedures\label{sec:changing_variables}}

In the previous sections we have presented the results for the
triple--differential
$b\to u l \bar{\nu}$ width to all orders in the large--$\beta_0$ limit.
We have chosen to describe the hadronic tensor in terms of the lightcone variables
$\alpha$ and $\beta$ and defined plus distributions with respect to $r=\beta/\alpha$.
Let us now shortly describe how the results can be used with other kinematic variables.
This is often useful for deriving analytic expressions for partially--integrated spectra, as done
for example in Ref.~\cite{DeFazio:1999sv} at the NLO level.

While the result for the virtual diagrams in a given regularization
is unique --- in the Borel regularization it is given by
Eqs. (\ref{V_I}) and (\ref{BV}) --- the infrared subtraction that renders
their perturbative expansion finite, namely Eqs.~(\ref{BV_reg}) and (\ref{BV_reg_}),
crucially depends on the corresponding real--emission terms that are put under the plus
prescription, \eq{B_i_small_beta},
and the variable ($r$) with respect to which the plus prescription is defined, see
Eqs.~(\ref{plus}) and (\ref{virt_sing}).

Let us first demonstrate how to use the results of the previous sections
in the same kinematic variables $\alpha$ and $\beta$, but with a different
infrared--subtraction convention.
Consider defining the plus distributions \emph{with respect to $\beta$};
instead of \eq{plus} we now write
\begin{equation}
\label{plus_beta}
\int_0^{\beta_0} d\beta F(\beta) \left[\frac{1}{\beta^{1+u}}\right]_{+}=\frac{F(0)}{u}\Big(1-\beta_0^{-u}\Big)
\,+\,\int_0^{\beta_0}d\beta
\Big(F(\beta)-F(0)\Big)\frac{1}{\beta^{1+u}},
\end{equation}
that corresponds to the replacement
\begin{equation}
\frac1{\beta^{1+u}} \longrightarrow \left[ \frac1{\beta^{1+u}} \right]_+-
\frac{\delta(\beta)}{u}.
\end{equation}

Taking the corresponding moments and applying (\ref{plus_beta}) we get ({\it cf.} Eq. (\ref{r_mom_def})):
\begin{eqnarray}
\label{beta_mom_def}
{\widetilde{\mathbb{W}}}_i^{(\beta)}(\alpha,\nu) &\equiv&
  \int_{0}^{1} {d\beta}
 \left(1- \beta \right)^{\nu-1}
{\mathbb{W}}_i(\alpha,\beta)
 \nonumber \\
&=& H_i^{(\beta)}(\alpha)\times  {{\rm
Sud}}^{(\beta)}(m_b,\alpha,\nu)+\Delta R_i^{(\beta)}(\alpha),
\end{eqnarray}
where the superscript $(\beta)$ is used to distinguish the current
definition from our default one, and
\begin{eqnarray}
\label{Sud_beta}
 \hspace*{-20pt}{{\rm Sud}}^{(\beta)}(m_b,\alpha,\nu)&=&
\exp\Bigg\{\frac{C_F}{\beta_0}\, \int_0^{\infty}\frac{du}{u}\,T(u)\,
 \left(\frac{\Lambda^2}{m_b^2}\right)^u
\Bigg[B_{\cal S}(u)
\left(\frac{\Gamma(\nu)\Gamma(-2u)}{\Gamma(\nu-2u)}+\frac{1}{2u}\right)
\nonumber\\ &&\hspace*{100pt}
 -\,\alpha^{-u}\,B_{\cal J}(u)\left(\frac{\Gamma(\nu)\Gamma(-u)}{\Gamma(\nu-u)}
 +\frac{1}{u}
\right)
 \Bigg]\Bigg\},
\end{eqnarray}
where the large--$\beta_0$ anomalous dimensions $B_{\cal S}(u)$ and $B_{\cal J}(u)$ are given in
(\ref{B_JS}). Note that the explicit $\alpha$ dependence in (\ref{Sud_beta}) is different from
(\ref{Sud}) owing to the different meaning of the moment variable $\nu$ compared to $N$.
The subtraction term $\left.B[V_0](\alpha,u)\right\vert_{\sing (\beta)}$, replacing (\ref{virt_sing}),
is therefore:
\begin{eqnarray}
\label{virt_sing_beta}
\left.B[V_0](\alpha,u)\right\vert_{\sing (\beta)}&\equiv&
\frac{1}{2u^2}\bigg[B_{\cal S}(u)-2\,\alpha^{-u}\,B_{\cal J}(u)
\bigg]
\\ \nonumber
&&\hspace*{-50pt}=\,
\frac{{\rm e}^{\frac53 u}}{2u^2} \,
\Bigg[(1-u)  -\,\alpha^{-u}\,\frac{\sin \pi u}{\pi u}\left(\frac{1}{1-u}+\frac{1}{1-u/2}
\right)\Bigg]
\\ \nonumber
&&\hspace*{-50pt}=\, -\frac{1}{2u^{2}}\,+\,\left(\ln(\alpha)-\frac{25}{12}\right)\frac{1}{u}
-\left(\frac{1}{2}\ln^2(\alpha)-\frac{29}{12}\ln(\alpha)-\frac{1}{6}\pi^2+\frac{245}{72}\right)
+{\cal O}(u).
\end{eqnarray}
Finally, using (\ref{virt_sing_beta}) in Eq.~(\ref{BV_reg}) we get
the corresponding infrared--subtracted version of the virtual terms that replaces (\ref{BV_reg_}) in
this alternative convention. The final results for the virtual terms, equivalent to
(\ref{BV_reg_expl}) immediately follow.

In a similar way one can consider the subtraction using other kinematic variables. A natural choice,
which was used in \cite{Aquila:2005hq} as well as in the original derivation of the singular terms
in~\cite{Gardi:2004ia}, is based on the invariant masses of the hadronic and the leptonic systems,
Eq.~(\ref{invariant_masses}). Let us define:
\begin{align}
\begin{split}
&1-\xi=\frac{p_j^2}{m_b^2}=\alpha\beta\,;\qquad \quad
z=\frac{q^2}{m_b^2}=(1-\alpha)(1-\beta)\qquad\\
& {\mathbb{W}}_i^{(\xi)}(z,\xi)={\mathbb{W}}_i(\alpha\,,\beta)
\left|
\frac{d(\alpha,\,\beta)}{d(z,\xi)}\right|=
\frac{1}{\alpha-\beta}{\mathbb{W}}_i(\alpha\,,\beta)
\end{split}
\end{align}
Here, infrared singularities are associated with the small jet mass limit $p_j^2\to 0$,
\begin{align}
\begin{split}
\label{B_i_small_xi}
 \left.B_{i}^{\SDG}
 (z,\xi,u)\right\vert_{\xi\to 1}&= \,
 \frac{V_i^{\rm LO}(\alpha=1-z)}{(1-\xi)^{1+u}}
 \frac{ {\rm e}^{\frac53 u}}{u}\,\times \,\\ & \hspace*{-45pt}
\left[(1-u)\left(\frac{1-\xi}{(1-z)^2}\right)^{-u}-
 \frac12\frac{\sin \pi u}{\pi u}\left(\frac{1}{1-u}+\frac{1}{1-u/2}
\right)
 \right]\,\times \Big(1+{\cal O}(1-\xi)\Big)\,,
\end{split}
\end{align}
so plus distributions are defined with respect to $\xi$ (i.e.
subtracting $\delta(1-\xi)$ terms) and the corresponding moments
are:
\begin{eqnarray}
\label{xi_mom_def} \widetilde{\mathbb{W}}_i^{(\xi)}(z,n) &\equiv&
  \int_{0}^{1} {d\xi}\,
 \xi^{n-1}\,
{\mathbb{W}}_i^{(\xi)}(z,\xi)
 \nonumber \\
&=& H_i^{(\xi)}(z)\times  {{\rm Sud}}^{(\xi)}(m_b,z,n)+\Delta
R_i^{(\xi)}(z),
\end{eqnarray}
with
\begin{eqnarray}
\label{Sud_xi}
 \hspace*{0pt}{{\rm Sud}}^{(\xi)}(m_b,z,n)&=&
\exp\Bigg\{\frac{C_F}{\beta_0}\, \int_0^{\infty}\frac{du}{u}\,T(u)\,
 \left(\frac{\Lambda^2}{m_b^2}\right)^u
\Bigg[B_{\cal S}(u) \,(1-z)^{2u}
\left(\frac{\Gamma(n)\Gamma(-2u)}{\Gamma(n-2u)}+\frac{1}{2u}\right)
\nonumber\\ &&\hspace*{100pt}
 -B_{\cal J}(u)\left(\frac{\Gamma(n)\Gamma(-u)}{\Gamma(n-u)}
 +\frac{1}{u}
\right)
 \Bigg]\Bigg\}.
\end{eqnarray}
Therefore, in these variables the subtraction term takes the form
\begin{eqnarray}
\label{virt_sing_xi}
\left.B[V_0](z,u)\right\vert_{\sing (\xi)}&\equiv&
\frac{1}{2u^2}\bigg[(1-z)^{2u}\,B_{\cal S}(u)\,-\,2\,B_{\cal J}(u)
\bigg]
\\ \nonumber
&&\hspace*{-90pt}=\,
\frac{{\rm e}^{\frac53 u}}{2u^2} \,
\Bigg[(1-z)^{2u}\,(1-u) \, -\,\frac{\sin \pi u}{\pi u}\left(\frac{1}{1-u}+\frac{1}{1-u/2}
\right)\Bigg]
\\ \nonumber
&&\hspace*{-90pt}=\, -\frac{1}{2u^{2}}\,+\,\left(\ln(1-z)-\frac{25}{12}\right)\frac{1}{u}
-\left(-\ln^2(1-z)-\frac{2}{3}\ln(1-z)-\frac{1}{6}\pi^2+\frac{245}{72}\right)
+{\cal O}(u).
\end{eqnarray}
As before, the corresponding infrared--subtracted virtual terms can be obtained using
(\ref{virt_sing_xi}) in Eqs.~(\ref{BV_reg}) with $\alpha\to 1-z$.

\section{Conclusions~\label{sec:Conc}}

We have computed the perturbative expansion of the triple
differential width in $b\to X_u l \bar{\nu}$, to all orders in the
large--$\beta_0$ limit. This is an important step in determining the
differential spectrum beyond the NLO.

Several independent partial
calculations have been done in the past that provided useful checks.
We find complete agreement with the following:
\begin{itemize}
\item{} The NLO calculation of the five structure functions (or the
fully differential width) in Ref.~\cite{DeFazio:1999sv}.
\item{} The NNLO result
of Ref.~\cite{vanRitbergen:1999gs}, Eq.~(\ref{total}) above, where
we could check the $\beta_0$ piece upon performing phase--space
integration according to (\ref{phase_space}).
\item{} The NNLO single--differential distribution with respect to
$p_j^+$, computed in the large--$\beta_0$ limit in
Ref.~\cite{Hoang:2005pj}, which we checked by defining the
subtraction procedure with respect to $\beta=p_j^+/m_b$ (see
Sec.~\ref{sec:changing_variables}) and then integrating over $x$ and
$\alpha$.
\item{} The singularity structure of the real--emission terms as
a function of the Borel variable in \cite{Gardi:2004ia}, and the
corresponding Sudakov exponent~\cite{Andersen:2005mj}.
\item{} The results of Ref.~\cite{Aquila:2005hq}, which have been
used here for the real--emission diagrams and computed by a
different method for the virtual ones.
\end{itemize}

As explained in the introduction, it is expected that the ${\cal
O}(\beta_0\alpha_s^2)$ contribution computed here constitutes the
bulk of the ${\cal O}(\alpha_s^2)$ correction. It therefore has an
immediate application in improving the calculation of partial
branching fractions used in the determination of $|V_{\rm ub}|$ from
inclusive measurements in the B--factories with a variety of kinematic cuts.

Although higher--order ${\cal O}(\beta_0^{n-1}\alpha_s^n)$
corrections, $n\geq 3$, may also be significant, we do not expect
that direct use of the single--dressed--gluon result by itself,
namely (\ref{Ri_borel}) and (\ref{Vi_reg_borel}) would yield a
viable description of the triple differential spectrum. Owing to the
$u\to \infty$ convergence constraint, the Borel integral of the
real--emission corrections (\ref{Ri_borel}) does not
exist\footnote{The implications have been studied in detail in
Ref.~\cite{Andersen:2006hr} in the context of $b\longrightarrow X_s
\gamma$; see Sec. 2.3 there.} for small~$p_j^+$, namely in the
Sudakov region. Better treatment of this region is achieved using
moment space~\cite{Gardi:2004ia,Andersen:2005mj}, where the
convergence constraint is replaced by infrared renormalons.
Moreover, in the Sudakov region the effect of \emph{multiple} soft
and collinear radiation is very important, and can be taken into
account by exponentiation (\ref{r_mom_def}), as done in
Ref.~\cite{Andersen:2005mj}. The running--coupling corrections
computed here can be used to improve the calculation of partial
branching fractions in the DGE approach of
Ref.~\cite{Andersen:2005mj} by incorporating the residual ${\cal
O}(\beta_0\alpha_s^2)$ correction that is not part of the Sudakov
factor into the matching coefficient.

Higher--order running--coupling corrections are also useful for
understanding the interplay between perturbative and
non-perturbative corrections, and for estimating the latter. Our
final results for the hadronic tensor, written as analytic functions
in the Borel plane, can be used to determine infrared renormalon
ambiguities, which are in turn indicative of the form and potential
magnitude of power corrections. In the virtual part of
(\ref{Vi_reg_borel}) with (\ref{BV_reg_expl}) one identifies
infrared renormalon singularities at integer and half integer values
of~$u$. In contrast, the differential real--emission contribution,
of (\ref{Ri_borel}) with (\ref{B_u}), does not present renormalon
singularities; as mentioned above these do show up in moment space
owing to the integration over $p_j^+$ near the singular $p_j^+\to 0$
limit. At the level of the Sudakov exponent this has already been
observed in Ref.~\cite{Gardi:2004ia} and been put to use in
Ref.~\cite{Andersen:2005mj}. The present results facilitate the
analysis of power corrections over the entire phase space.

Let us end with a brief comment on the technical tools used
in this paper, which made it possible to derive analytic expressions
for the Borel transform. We
exploited two different techniques for the calculation of Feynman
diagrams with a single dressed gluon:
\begin{itemize}
\item{}
The real--emission diagrams were computed in
Ref.~\cite{Aquila:2005hq} using the dispersive approach, where the
gluon in the final state is assigned a fixed virtuality, which is
then used in a dispersive integral with the time--like discontinuity
of the coupling. Here we converted the result into a Borel
representation and derived analytic expressions for the Borel
function.
\item{} The virtual diagrams were computed here directly in terms of
the Borel variable, by modifying the gluon propagator according to
\eq{prop}, and then preforming the momentum integral.
\end{itemize}
Having brought the results of both real and virtual diagrams with a
single dressed gluon to a common regularization, we could directly
perform \emph{an all--order infrared subtraction}. In this novel
approach the Borel variable has a double r\^ole: on the one hand it
serves as an infrared regulator for logarithmic singularities --- a
double pole at $u=0$, in full analogy with dimensional
regularization --- and on the other, it serves as a conjugate to
$\ln m_b^2/\Lambda^2$, or the inverse of the coupling constant,
allowing for all--order resummation of running--coupling
corrections.

\subsection*{Acknowledgements}
The work of PG is supported in part by MIUR under contract 2004021808-009.

\appendix

\section{The scheme--invariant Borel
transform\label{sec:scheme_invariant_Borel}}

In this paper, e.g. in Eqs.~(\ref{Ri_borel}) and (\ref{Vi_reg_borel}), we use
the scheme--invariant~\cite{Grunberg:1992hf} Borel representation
where $T(u)$ is defined as the inverse Laplace transform of the coupling:
\begin{equation}
\frac{\beta_0\alpha_s(\mu)}{\pi}= \int_0^{\infty}{du} \,T(u)\,
\left(\frac{\Lambda^2}{\mu^2}\right)^u.
\end{equation}
Here the Borel variable~$u$ is the Laplace conjugate to the
logarithm $\ln \mu^2/\Lambda^2$, rather than to the inverse of the
coupling constant, which is used in the standard Borel transform. In
this way it is possible to promote the calculation performed in the
large--$\beta_0$ limit to include running--coupling terms that are
associated with subleading corrections to the $\beta$ function, in
particular $\beta_1$. This was used in various applications, see
e.g. Ref.~\cite{Gardi:2002xm}.

In the strict large--$\beta_0$ limit, one resums the terms
$\beta_0^{n-1}\alpha_s^n$ to any $n$. In this case ${\mbox
T(u)\equiv 1}$, and \eq{Ri_borel} (or \eq{Vi_reg_borel} for the
virtual terms)
reduces to the standard Borel
transform, with respect to
$A_{\MSbar}(\mu)={\beta_0\alpha_s^{\MSbar}(\mu)}/{\pi}$, namely
\begin{eqnarray}
\label{Standard_borel} R_i^{\rm large \,\,\beta_0}(\alpha,\beta)&=&
\frac{C_F}{\beta_0}
\int_0^{\infty}du\,\exp\Big\{-u/A_{\MSbar}(\mu)\Big\}\,
\left(\frac{\mu^2}{m_b^2}\right)^u\,B_i^{\SDG}(\alpha,\beta,u)\\
&=& C_F\,\bigg[c_i^{(1)}(\alpha,\beta)\,\frac{\alpha_s(m_b)}{\pi}+
c_i^{(2)}(\alpha,\beta)\,\beta_0\left(\frac{\alpha_s(m_b)}{\pi}\right)^2
+\cdots\bigg],
\nonumber
\end{eqnarray}
with the relation:
\begin{equation}
\label{expanding_Borel}
B_i^{\SDG}(\alpha,\beta,u)=\sum_{n=1}^{\infty}
c_i^{(n)}(\alpha,\beta) \,\frac{u^n}{n!}.
\end{equation}
Upon using the scheme--invariant formulation as in \eq{Ri_borel},
it is straightforward to include $\beta_1$
effects in the running of the coupling by introducing $T(u)$ that
corresponds to the Laplace transform of the two--loop coupling (or
the 't~Hooft--scheme coupling):
\begin{eqnarray}
\label{tHooft_coupling}
A(\mu)&=&\frac{\beta_0\alpha_s^{\tHooft}(\mu)}{\pi}=
\int_0^{\infty}{du} \,T(u)\, \left(\frac{\Lambda^2}{\mu^2}\right)^u;
\qquad \qquad \frac{dA}{d\ln \mu^2} =-A^2(1+\delta A),
\nonumber \\
T(u)&=&\frac{(u\delta)^{u\delta}{\rm
e}^{-u\delta}}{\Gamma(1+u\delta)};\qquad \qquad \ln
(\mu^2/\Lambda^2)=\frac{1}{A}-\delta\ln\left(1+\frac{1}{\delta
A}\right),
\end{eqnarray}
with $\delta\equiv \beta_1/\beta_0^2$. Upon expanding \eq{Ri_borel}, or
\eq{Vi_reg_borel}, with $T(u)$ of (\ref{tHooft_coupling}) one obtains,
in addition to
the large--$\beta_0$ terms, also $\beta_1 \beta_0^{n-3}\alpha_s^n$
terms, etc.

\section{The functions of the lightcone variables
entering the Borel transform\label{sec:DST_explicit_expressions}}

Here we give explicit results for the rational functions of the lightcone variables,
 $D_{i,j}(\alpha,\beta)$, $S_{i,j}(\alpha,\beta)$ and $T_{i,j}(\alpha,\beta)$
 entering the Borel function of \eq{DST}.
Recall that the first index here, $i=1$ to 5, corresponds to the
structure function, while the second to the location of the
singularity on the positive real axis in the Borel plane, $u=j$ with
$j=0$ to $2$. There are a few relations among them, valid for any $i$:
\begin{align}
\begin{split}
D_{i,0}(\alpha,\beta) = (\alpha+\beta-2\, \alpha\,\beta)\, S_{i,0}(\alpha,\beta)
\end{split}
\end{align}
\begin{align}
\begin{split}
\widetilde{D}_{i,1}(\alpha,\beta) = -(\alpha+\beta)\,
\widetilde{S}_{i,1}(\alpha,\beta)
\end{split}
\end{align}
\begin{align}
\begin{split}
D_{i,2}(\alpha,\beta) = -\frac{\alpha^3(1-\beta)+\beta^3(1-\alpha)}{
\alpha^2+\beta^2+\alpha\beta(1-\alpha-\beta)} \,S_{i,2}(\alpha,\beta)
\end{split}
\end{align}

The remaining functions are:
\begin{align}
\begin{split}
{{S}_{1,2}}(\alpha,\beta) =& {\displaystyle \frac {\,( - \alpha  - \beta  + \alpha ^{2} + \beta ^{2} + 3
\,\alpha \,\beta  + 3\,\alpha ^{2}\,\beta ^{2} - 3\,\alpha ^{2}\,
\beta  - 3\,\alpha \,\beta ^{2})}{2\,( - 1 + \beta )\,( - 1 + \alpha
)\,( - \beta  - \alpha  + \alpha \,\beta )}}\times\\
& ( - \alpha ^{2}
 + \alpha ^{2}\,\beta  - \alpha \,\beta  + \alpha \,\beta ^{2} -
\beta ^{2})
\end{split}
\end{align}
\begin{align}
\begin{split}
{\widetilde{S}_{1,1}}(\alpha,\beta) =& ( - \alpha ^{2} + 2\,\alpha ^{2}\, \beta
^{3} + 2\,\alpha ^{3} - \beta ^{2} - 3\,\alpha ^{3}\,\beta
 + 2\,\beta ^{3} + 2\,\alpha ^{2}\,\beta  + 2\,\alpha ^{3}\,\beta
 ^{2} - 4\,\alpha ^{2}\,\beta ^{2} \\&
\mbox{}  + 2\,\alpha \,\beta - 3\,\alpha \,\beta ^{3} + 2\,\alpha \,\beta ^{2})/(2\,(
 - \beta  - \alpha  + \alpha \,\beta ))
\end{split}
\end{align}
\begin{align}
\begin{split}
{{S}_{1,1}}(\alpha,\beta) =&  - \beta \,\alpha ( - 11\,\alpha ^{3}\, \beta  +
5\,\alpha ^{3} + 6\,\alpha ^{3}\,\beta ^{2} + 31\,\alpha
 ^{2}\,\beta  - 15\,\alpha ^{2} - 22\,\alpha ^{2}\,\beta ^{2} + 6
\,\alpha ^{2}\,\beta ^{3}  \\& \mbox{}- 34\,\alpha \,\beta  +
12\,\alpha  - 11\,\alpha \,\beta ^{3} + 31\,\alpha \, \beta ^{2} -
15\,\beta ^{2} + 5\,\beta ^{3} + 12\,\beta )/ \\&
\left(4\,(
 - 1 + \beta )\,( - 1 + \alpha )( - \beta  - \alpha  + \alpha \,\beta )\right)
\end{split}
\end{align}
\begin{align}
\begin{split}
{{S}_{1,0}}(\alpha,\beta) =& {\displaystyle \frac {2\,\beta ^{4} - 2 \,\alpha
\,\beta ^{4} + \alpha ^{2}\,\beta ^{4} - 4\,\alpha ^{2} \,\beta ^{3}
- 4\,\alpha ^{2}\,\beta ^{2} - 4\,\alpha ^{3}\,\beta
 ^{2} + \alpha ^{4}\,\beta ^{2} - 2\,\alpha ^{4}\,\beta  + 2\,
\alpha ^{4}}{4\,\alpha \,\beta \,( - \beta  - \alpha  + \alpha \,
\beta )}}
\end{split}
\end{align}
\begin{align}
\begin{split}
{{D}_{1,1}}(\alpha,\beta) =& (3\,\alpha ^{4}\,\beta ^{3} + 13\, \alpha
^{4}\,\beta  + 3\,\alpha ^{3}\,\beta ^{4} - 16\,\alpha ^{2 }\,\beta
^{4} + 13\,\alpha \,\beta ^{4} - 16\,\alpha ^{4}\,\beta ^{2} +
2\,\alpha ^{3}\,\beta ^{5} + 2\,\alpha ^{5}\,\beta ^{3}
 \\&
\mbox{} + \alpha \,\beta ^{5} + \alpha ^{5}\,\beta  - 3\,\alpha
^{5}\,\beta ^{2} - 3\,\alpha ^{2}\,\beta ^{5} + 4\,\alpha ^{3} +
11\,\alpha ^{3}\,\beta ^{2} - 8\,\alpha ^{3}\,\beta  + 11\,\alpha
 ^{2}\,\beta ^{3}  \\&
\mbox{} + 8\,\alpha ^{2}\,\beta ^{2}- 4\,\alpha ^{2}\,\beta  - 8\,\alpha \,\beta ^{3} - 4\,
\alpha \,\beta ^{2} + 4\,\beta ^{3} - 10\,\alpha ^{3}\,\beta ^{3}
 + 4\,\alpha ^{4}\,\beta ^{4} - 4\,\alpha ^{4} \\&
- 4\,\beta ^{4})/ (4\,( - 1 + \beta ) ( - 1 + \alpha )\,( - \beta  - \alpha  +
\alpha \,\beta ))
\end{split}
\end{align}
\begin{align}
\begin{split}
{{T}_{1,2}}(\alpha,\beta) =& - (6\,\alpha ^{4}\,\beta ^{3} - 8\, \alpha
^{4}\,\beta ^{2} + \alpha ^{4}\,\beta  + \alpha ^{4} + 9\, \alpha
^{3}\,\beta ^{2} - 2\,\alpha ^{3}\,\beta  - \alpha ^{3} + 6\,\alpha
^{3}\,\beta ^{4} - 8\,\alpha ^{3}\,\beta ^{3} \\& \mbox{} -
6\,\alpha ^{2}\,\beta ^{2} + \alpha ^{2}\,\beta  + 9\, \alpha
^{2}\,\beta ^{3} - 8\,\alpha ^{2}\,\beta ^{4} + \alpha \, \beta ^{4}
- 2\,\alpha \,\beta ^{3} + \alpha \,\beta ^{2} - \beta
 ^{3} + \beta ^{4})/\\&
(4\,( - 1 + \beta ) ( - 1 + \alpha )\,\alpha \,\beta )
\end{split}
\end{align}
\begin{align}
\begin{split}
{{T}_{1,1}}(\alpha,\beta) =& (4\,\alpha ^{4}\,\beta ^{3} - 8\,\alpha
 ^{4}\,\beta ^{2} + 5\,\alpha ^{4}\,\beta  - \alpha ^{4} - 12\,
\alpha ^{3}\,\beta ^{3} + 13\,\alpha ^{3}\,\beta ^{2} - 4\,\alpha
 ^{3}\,\beta  + \alpha ^{3} + 4\,\alpha ^{3}\,\beta ^{4} \\&
\mbox{} - 8\,\alpha ^{2}\,\beta ^{4} - 6\,\alpha ^{2}\,\beta ^{2}
 - \alpha ^{2}\,\beta  + 13\,\alpha ^{2}\,\beta ^{3} + 5\,\alpha
\,\beta ^{4} - 4\,\alpha \,\beta ^{3} - \alpha \,\beta ^{2} + \beta
^{3} - \beta ^{4})/(2 \\& ( - 1 + \beta )\,( - 1 + \alpha )\,\alpha
\,\beta )
\end{split}
\end{align}
\begin{align}
\begin{split}
{{T}_{1,0}}(\alpha,\beta) =& {\displaystyle \frac { - 7\,\alpha \,\beta (\alpha+\beta)  
- 10\,\alpha ^{3}\,\beta  - 10 \,\alpha
\,\beta ^{3} + 2\,\alpha ^{2}\,\beta ^{3} + 7\,\beta ^{3 } -
4\,\alpha ^{2}\,\beta ^{2} + 2\,\alpha ^{3}\,\beta ^{2} + 7\, \alpha
^{3}}{4\,\alpha \,\beta }}
\end{split}
\end{align}
\begin{align}
\begin{split}
{{S}_{2,2}}(\alpha,\beta) =&  - 2\,(9\,\alpha ^{2}\,\beta ^{2} - 9\, \alpha
^{2}\,\beta  + \alpha ^{2} - 9\,\alpha \,\beta ^{2} + 13\, \alpha
\,\beta  - 3\,\alpha  + \beta ^{2} - 3\,\beta ) \\& ( - \alpha ^{2}
+ \alpha ^{2}\,\beta  - \alpha \,\beta  + \alpha \,\beta ^{2} -
\beta ^{2})/( - \beta  - \alpha  + \alpha \,\beta )
\end{split}
\end{align}
\begin{align}
\begin{split}
{\widetilde{S}_{2,1}}(\alpha,\beta) =&  - 2(9\,\alpha \,\beta ^{2} - 2\,\beta
 ^{4} + 39\,\alpha ^{2}\,\beta ^{3} + 6\,\alpha ^{4}\,\beta ^{3}
 + 39\,\alpha ^{3}\,\beta ^{2} - 34\,\alpha ^{2}\,\beta ^{2} + 9
\,\alpha ^{2}\,\beta  - 2\,\alpha ^{4} \\& \mbox{} + 3\,\alpha ^{3}
- 20\,\alpha ^{3}\,\beta  - 13\,\alpha ^{4}\,\beta ^{2} - 20\,\alpha
\,\beta ^{3} + 9\,\alpha \,\beta ^{ 4} + 9\,\alpha ^{4}\,\beta  -
28\,\alpha ^{3}\,\beta ^{3} - 13\, \alpha ^{2}\,\beta ^{4} \\&
\mbox{} + 6\,\alpha ^{3}\,\beta ^{4} + 3\,\beta ^{3})/( - \beta
 - \alpha  + \alpha \,\beta )
\end{split}
\end{align}
\begin{align}
\begin{split}
{{S}_{2,1}}(\alpha,\beta) =& ( - 35\,\alpha ^{4}\,\beta ^{2} + 18\, \alpha
^{4}\,\beta ^{3} + 17\,\alpha ^{4}\,\beta  + 91\,\alpha ^{ 3}\,\beta
^{2} - 62\,\alpha ^{3}\,\beta ^{3} + 6\,\alpha ^{3} - 52\,\alpha
^{3}\,\beta  \\& \mbox{} + 18\,\alpha ^{3}\,\beta ^{4} - 35\,\alpha
^{2}\,\beta ^{ 4} - 88\,\alpha ^{2}\,\beta ^{2} + 30\,\alpha
^{2}\,\beta  + 91\, \alpha ^{2}\,\beta ^{3} + 17\,\alpha \,\beta
^{4} - 52\,\alpha \, \beta ^{3} \\& \mbox{} + 30\,\alpha \,\beta
^{2} + 6\,\beta ^{3})/( - \beta  - \alpha  + \alpha \,\beta )
\end{split}
\end{align}
\begin{align}
\begin{split}
{{S}_{2,0}}(\alpha,\beta) =&  - ( - 21\,\alpha ^{4}\,\beta ^{3} - 2 \,\alpha
^{4}\,\beta  - 21\,\alpha ^{3}\,\beta ^{4} + 18\,\alpha ^{2}\,\beta
^{4} - 2\,\alpha \,\beta ^{4} + 34\,\alpha ^{3}\, \beta ^{3} +
4\,\alpha ^{4}\,\beta ^{4} \\& \mbox{} - 14\,\alpha ^{3}\,\beta ^{2}
- 14\,\alpha ^{2}\,\beta ^{ 3} + \alpha ^{5}\,\beta ^{3} + 4\,\alpha
\,\beta ^{5} + 18\, \alpha ^{4}\,\beta ^{2} + \alpha ^{3}\,\beta
^{5} - 3\,\alpha ^{2 }\,\beta ^{5} - 2\,\alpha ^{5} \\& \mbox{} +
4\,\alpha ^{5}\,\beta  - 3\,\alpha ^{5}\,\beta ^{2} - 2 \,\beta
^{5})/(\alpha \,\beta \,( - \beta  - \alpha  + \alpha \, \beta ))
\end{split}
\end{align}

\begin{align}
\begin{split}
{{D}_{2,1}}(\alpha,\beta) =&  - ( - 12\,\alpha \,\beta ^{3} - 12\, \alpha
^{3}\,\beta  + 27\,\alpha ^{3}\,\beta ^{4} + 14\,\alpha ^{ 5}\,\beta
^{3} + 14\,\alpha ^{3}\,\beta ^{5} - 27\,\alpha ^{2}\, \beta ^{5} -
27\,\alpha ^{5}\,\beta ^{2} \\& \mbox{} - 4\,\alpha \,\beta ^{4} -
20\,\alpha ^{2}\,\beta ^{4} - 20\,\alpha ^{4}\,\beta ^{2} + 2\,\beta
^{4} + 2\,\alpha ^{4} - 4 \,\alpha ^{4}\,\beta  + 27\,\alpha
^{4}\,\beta ^{3} + 76\,\alpha ^{3}\,\beta ^{2} \\& \mbox{} +
76\,\alpha ^{2}\,\beta ^{3} + 13\,\alpha \,\beta ^{5}
 + 13\,\alpha ^{5}\,\beta  - 106\,\alpha ^{3}\,\beta ^{3} - 4\,
\alpha ^{4}\,\beta ^{4}\\& - 28\,\alpha ^{2}\,\beta ^{2})/( - \beta
 - \alpha  + \alpha \,\beta )
\end{split}
\end{align}
\begin{align}
\begin{split}
{{T}_{2,2}}(\alpha,\beta) =& (18\,\alpha ^{4}\,\beta ^{3} - 24\, \alpha
^{4}\,\beta ^{2} + 5\,\alpha ^{4}\,\beta  + \alpha ^{4} + 25\,\alpha
^{3}\,\beta ^{2} - 14\,\alpha ^{3}\,\beta  + 18\, \alpha ^{3}\,\beta
^{4} - 24\,\alpha ^{3}\,\beta ^{3} \\& \mbox{} + 2\,\alpha
^{2}\,\beta ^{2} + 25\,\alpha ^{2}\,\beta ^{3 } - 24\,\alpha
^{2}\,\beta ^{4} + 5\,\alpha \,\beta ^{4} - 14\, \alpha \,\beta ^{3}
+ \beta ^{4})/(\alpha \,\beta )
\end{split}
\end{align}
\begin{align}
\begin{split}
{{T}_{2,1}}(\alpha,\beta) =&  - 2\,( - \beta  + 2\,\alpha \,\beta
 - \alpha )\,( - 7\,\alpha ^{3}\,\beta  + 6\,\alpha ^{3}\,\beta ^{2} + \alpha
^{3} + 6\,\alpha ^{2}\,\beta ^{3} - 16\,\alpha ^{2}\,\beta ^{2}
\\& + 11\,\alpha ^{2}\,\beta  - 7\,\alpha \,\beta ^{3} + 11\,\alpha
\,\beta ^{2}   + \beta ^{3})/(\alpha \beta )
\end{split}
\end{align}
\begin{align}
\begin{split}
{{T}_{2,0}}(\alpha,\beta) =&  - ( - 7\,\beta ^{4} - 20\,\alpha ^{4} \,\beta
^{2} - 10\,\alpha ^{3}\,\beta  - 20\,\alpha ^{2}\,\beta ^{4} -
10\,\alpha \,\beta ^{3} - 56\,\alpha ^{3}\,\beta ^{3} + 21 \,\alpha
^{4}\,\beta  - 7\,\alpha ^{4} \\& \mbox{} - 38\,\alpha ^{2}\,\beta
^{2} + 57\,\alpha ^{3}\,\beta ^{ 2} + 21\,\alpha \,\beta ^{4} +
57\,\alpha ^{2}\,\beta ^{3} + 6\, \alpha ^{4}\,\beta ^{3} +
6\,\alpha ^{3}\,\beta ^{4})/(\alpha \, \beta )
\end{split}
\end{align}
\begin{align}
\begin{split}
{{S}_{3,2}}(\alpha,\beta) =&  - {\displaystyle \frac { - \alpha ^{2}
 + \alpha ^{2}\,\beta  - \alpha \,\beta  + \alpha \,\beta ^{2} -
\beta ^{2}}{ - \beta  - \alpha  + \alpha \,\beta }}
\end{split}
\end{align}
\begin{align}
\begin{split}
{\widetilde{S}_{3,1}}(\alpha,\beta) =& {\displaystyle \frac { - 3\,\alpha \,
\beta ^{2} + 2\,\beta ^{2} + 2\,\alpha \,\beta  + 2\,\alpha ^{2}
\,\beta ^{2} - 3\,\alpha ^{2}\,\beta  - \beta  + 2\,\alpha ^{2}
 - \alpha }{ - \beta  - \alpha  + \alpha \,\beta }}
\end{split}
\end{align}
\begin{align}
\begin{split}
{{S}_{3,1}}(\alpha,\beta) =&  - {\displaystyle \frac {\beta \,\alpha
 \,(2\,\alpha \,\beta  - 3\,\alpha  + 4 - 3\,\beta )}{2\,( -
\beta  - \alpha  + \alpha \,\beta )}}
\end{split}
\end{align}
\begin{align}
\begin{split}
{{S}_{3,0}}(\alpha,\beta) =& {\displaystyle \frac { - 2\,\alpha \, \beta ^{3} +
\alpha ^{3}\,\beta ^{2} + \alpha ^{2}\,\beta ^{3} + 2\,\alpha ^{3} +
2\,\beta ^{3} + 2\,\alpha ^{2}\,\beta  - 6\, \alpha ^{2}\,\beta ^{2}
- 2\,\alpha ^{3}\,\beta  + 2\,\alpha \, \beta ^{2}}{2\,\alpha
\,\beta \,( - \beta  - \alpha  + \alpha \, \beta )}}
\end{split}
\end{align}
\begin{align}
\begin{split}
{{D}_{3,1}}(\alpha,\beta) =&  - {\displaystyle \frac {(\alpha ^{2}\, \beta  +
\alpha \,\beta ^{2} - 8\,\alpha \,\beta  + 4\,\alpha  + 4\,\beta
)\,( - \beta  + 2\,\alpha \,\beta  - \alpha )}{2\,( - \beta - \alpha
+ \alpha \,\beta )}}
\end{split}
\end{align}
\begin{align}
\begin{split}
{{T}_{3,2}}(\alpha,\beta) =& {\displaystyle \frac {2\,\alpha ^{2}\, \beta  +
\alpha ^{2} - 2\,\alpha \,\beta  + 2\,\alpha \,\beta ^{2 } + \beta
^{2}}{2\,\alpha \,\beta }}
\end{split}
\end{align}
\begin{align}
\begin{split}
{{T}_{3,1}}(\alpha,\beta) =& {\displaystyle \frac {( - \beta  + 2\, \alpha
\,\beta  - \alpha )^{2}}{\alpha \,\beta }}
\end{split}
\end{align}
\begin{align}
\begin{split}
{{T}_{3,0}}(\alpha,\beta) =& {\displaystyle \frac {7\,\alpha ^{2} + 7\,\beta
^{2} + 4\,\alpha ^{2}\,\beta ^{2} - 10\,\alpha \,\beta ^{2} +
2\,\alpha \,\beta  - 10\,\alpha ^{2}\,\beta }{2\,\alpha \, \beta }}
\end{split}
\end{align}
\begin{align}
\begin{split}
{{S}_{4,2}}(\alpha,\beta) =&  - ( - \alpha ^{2} + \alpha ^{2}\,\beta
  - \alpha \,\beta  + \alpha \,\beta ^{2} - \beta ^{2})
  (4\,\alpha ^{2}- 6\,
\alpha  - 6\,\beta   + 4\,\beta ^{2} + 22\,\alpha
 \,\beta  + 12\,\alpha ^{2}\,\beta ^{2} \\&
\mbox{} - 18\,\alpha ^{2}\,\beta  - 18\,\alpha \,\beta ^{2} + 3\,
\alpha \,\beta ^{3} + 3\,\alpha ^{3}\,\beta )/(( - 1 + \beta )\,(
 - 1 + \alpha )\,( - \beta  - \alpha  + \alpha \,\beta ))
\end{split}
\end{align}
\begin{align}
\begin{split}
{\widetilde{S}_{4,1}}(\alpha,\beta) =&  - 2(2\,\beta ^{3} + 5\,\alpha ^{2}\,
\beta ^{3} + 2\,\alpha ^{3} + 10\,\alpha ^{2}\,\beta  - 8\,\alpha
 \,\beta ^{3} + 5\,\alpha ^{3}\,\beta ^{2} - 14\,\alpha ^{2}\,
\beta ^{2} + 10\,\alpha \,\beta ^{2} \\& \mbox{} + \alpha \,\beta
^{4} + \alpha ^{4}\,\beta  - 8\,\alpha ^{3}\,\beta )/( - \beta  -
\alpha  + \alpha \,\beta )
\end{split}
\end{align}
\begin{align}
\begin{split}
{{S}_{4,1}}(\alpha,\beta) =& (195\,\alpha ^{4}\,\beta ^{3} + 78\, \alpha
^{4}\,\beta  + 195\,\alpha ^{3}\,\beta ^{4} - 187\,\alpha
^{2}\,\beta ^{4} + 78\,\alpha \,\beta ^{4} - 187\,\alpha ^{4}\,
\beta ^{2} - 51\,\alpha ^{3}\,\beta ^{5} \\& \mbox{} - 51\,\alpha
^{5}\,\beta ^{3} - 28\,\alpha \,\beta ^{5}
 - 28\,\alpha ^{5}\,\beta  + 60\,\alpha ^{5}\,\beta ^{2} + 15\,
\alpha ^{5}\,\beta ^{4} + 15\,\alpha ^{4}\,\beta ^{5} + 60\, \alpha
^{2}\,\beta ^{5}  \\& \mbox{} + 6\,\alpha ^{3}+ 3\,\alpha
^{3}\,\beta ^{6} + 3\,\alpha ^{6}\,\beta ^{3}
 + 3\,\alpha ^{6}\,\beta  + 3\,\alpha \,\beta ^{6} - 6\,\alpha ^{
2}\,\beta ^{6} - 6\,\alpha ^{6}\,\beta ^{2} + 4\,\beta ^{5} + 4\,
\alpha ^{5} \\& \mbox{} + 254\,\alpha ^{3}\,\beta ^{2} - 84\,\alpha
^{3}\,\beta
 + 254\,\alpha ^{2}\,\beta ^{3} - 148\,\alpha ^{2}\,\beta ^{2} +
30\,\alpha ^{2}\,\beta  - 84\,\alpha \,\beta ^{3} + 30\,\alpha \,
\beta ^{2} \\& \mbox{} + 6\,\beta ^{3} - 326\,\alpha ^{3}\,\beta
^{3} - 90\, \alpha ^{4}\,\beta ^{4} - 10\,\alpha ^{4} - 10\,\beta
^{4})
 \left/ {\vrule height0.43em width0em depth0.43em} \right. \!
 \! (( - \beta  - \alpha  + \alpha \,\beta )\,( - 1 + \alpha )^{2
} \\& ( - 1 + \beta )^{2})
\end{split}
\end{align}
\begin{align}
\begin{split}
{{S}_{4,0}}(\alpha,\beta) =&  - {\displaystyle \frac {2\,( - 2\, \alpha ^{3} +
\alpha \,\beta ^{3} - 2\,\beta ^{3} - 7\,\alpha ^{2 }\,\beta +
\alpha ^{3}\,\beta  + \alpha ^{2}\,\beta ^{2} - 7\, \alpha \,\beta
^{2})}{ - \beta  - \alpha  + \alpha \,\beta }}
\end{split}
\end{align}
\begin{align}
\begin{split}
{{D}_{4,1}}(\alpha,\beta) =&  - (3\,\alpha ^{7}\,\beta  + 207\, \alpha
^{4}\,\beta ^{3} + 6\,\alpha ^{4}\,\beta  + 207\,\alpha ^{ 3}\,\beta
^{4} - 116\,\alpha ^{2}\,\beta ^{4} + 6\,\alpha \,\beta
 ^{4} - 116\,\alpha ^{4}\,\beta ^{2} \\&
\mbox{} - 29\,\alpha ^{3}\,\beta ^{5} - 29\,\alpha ^{5}\,\beta ^{ 3}
+ 22\,\alpha \,\beta ^{5} + 22\,\alpha ^{5}\,\beta  + 5\, \alpha
^{5}\,\beta ^{2} + 5\,\alpha ^{5}\,\beta ^{4} + 5\,\alpha
^{4}\,\beta ^{5}  \\& \mbox{} + 5\,\alpha ^{2}\,\beta ^{5}+
3\,\alpha ^{3}\,\beta ^{7} + 6\,\alpha ^{6}\,\beta ^{4}
 + 6\,\alpha ^{4}\,\beta ^{6} + 6\,\alpha ^{5}\,\beta ^{5} + 3\,
\alpha ^{7}\,\beta ^{3} - 23\,\alpha ^{3}\,\beta ^{6}  \\&
 \mbox{}- 23\, \alpha
^{6}\,\beta ^{3} - 20\,\alpha ^{6}\,\beta  - 20\,\alpha
\,\beta ^{6} + 33\,\alpha ^{2}\,\beta ^{6}
 + 33\,\alpha ^{6}\,\beta ^{2} + 4\,\alpha ^{6} + 4\,\beta ^{6}
  \\& \mbox{} + 3\,\alpha \,\beta ^{7} - 6\,\alpha ^{2}\,\beta ^{7} - 10\,
\beta ^{5} - 10\,\alpha ^{5}- 6\,\alpha ^{7}\,\beta
^{2} + 124\,\alpha ^{3}\,\beta ^{ 2} - 12\,\alpha ^{3}\,\beta \\& +
124\,\alpha ^{2}\,\beta ^{3} - 36 \,\alpha ^{2}\,\beta ^{2} -
12\,\alpha \,\beta ^{3} - 276\,\alpha
 ^{3}\,\beta ^{3} \\&
\mbox{} - 110\,\alpha ^{4}\,\beta ^{4} + 6\,\alpha ^{4} + 6\, \beta
^{4}) \left/ {\vrule height0.43em width0em depth0.43em}
 \right. \!  \! (( - \beta  - \alpha  + \alpha \,\beta )\,( - 1
 + \alpha )^{2}\,( - 1 + \beta )^{2})
\end{split}
\end{align}
\begin{align}
\begin{split}
{{T}_{4,2}}(\alpha,\beta) =& (3\,\alpha ^{4}\,\beta  - \alpha ^{4}
 + 15\,\alpha ^{3}\,\beta ^{2} - 20\,\alpha ^{3}\,\beta  + 7\,
\alpha ^{3} - 30\,\alpha ^{2}\,\beta ^{2} + 23\,\alpha ^{2}\, \beta
- 10\,\alpha ^{2} + 15\,\alpha ^{2}\,\beta ^{3} \\& \mbox{} +
3\,\alpha \,\beta ^{4} - 20\,\alpha \,\beta ^{3} - 4\, \alpha
\,\beta  + 23\,\alpha \,\beta ^{2} - \beta ^{4} + 7\,\beta
 ^{3} - 10\,\beta ^{2})/(( - 1 + \beta )\,( - 1 + \alpha ))
\end{split}
\end{align}
\begin{align}
\begin{split}
{{T}_{4,1}}(\alpha,\beta) =&  - 2(10\,\alpha ^{4}\,\beta ^{3} + 35\, \alpha
^{4}\,\beta  + 10\,\alpha ^{3}\,\beta ^{4} - 34\,\alpha ^{ 2}\,\beta
^{4} + 35\,\alpha \,\beta ^{4} - 34\,\alpha ^{4}\,\beta
 ^{2} - 3\,\alpha \,\beta ^{5} \\&
\mbox{} - 3\,\alpha ^{5}\,\beta  + 2\,\alpha ^{5}\,\beta ^{2} + 2
\,\alpha ^{2}\,\beta ^{5} + 23\,\alpha ^{3} + \beta ^{5} + \alpha
 ^{5} - 24\,\alpha \,\beta  - 12\,\beta ^{2} - 12\,\alpha ^{2}
 \\&
\mbox{} + 114\,\alpha ^{3}\,\beta ^{2} - 86\,\alpha ^{3}\,\beta
 + 114\,\alpha ^{2}\,\beta ^{3} - 152\,\alpha ^{2}\,\beta ^{2} +
79\,\alpha ^{2}\,\beta  - 86\,\alpha \,\beta ^{3} + 79\,\alpha \,
\beta ^{2} \\& \mbox{} + 23\,\beta ^{3} - 58\,\alpha ^{3}\,\beta
^{3} - 12\, \alpha ^{4} - 12\,\beta ^{4}) \left/ {\vrule
height0.43em width0em depth0.43em} \right. \!  \! (( - 1 + \alpha
 )^{2}\,( - 1 + \beta )^{2})
\end{split}
\end{align}
\begin{align}
\begin{split}
{{T}_{4,0}}(\alpha,\beta) =& 22\,\beta ^{2} + 22\,\alpha ^{2} - 5\, \alpha
^{2}\,\beta  - 5\,\alpha \,\beta ^{2} + 28\,\alpha \,\beta
  - \alpha ^{3} - \beta ^{3}
\end{split}
\end{align}
\begin{align}
\begin{split}
{{S}_{5,2}}(\alpha,\beta) =&  - ( - \alpha ^{2} + \alpha ^{2}\,\beta
  - \alpha \,\beta  + \alpha \,\beta ^{2} - \beta ^{2})(6\,\alpha
  + 6\,\beta  - 6\,\alpha ^{2} - 6\,\beta ^{2} - 30\,\alpha \,
\beta  - 36\,\alpha ^{2}\,\beta ^{2} \\& \mbox{} + 32\,\alpha
^{2}\,\beta  + 32\,\alpha \,\beta ^{2} + \beta ^{3} + \alpha ^{3} +
9\,\alpha ^{2}\,\beta ^{3} + 9\,\alpha
 ^{3}\,\beta ^{2} - 9\,\alpha \,\beta ^{3} - 9\,\alpha ^{3}\,
\beta )/(( - 1 + \beta ) \\& ( - 1 + \alpha )\,( - \beta  - \alpha +
\alpha \,\beta ))
\end{split}
\end{align}
\begin{align}
\begin{split}
{\widetilde{S}_{5,1}}(\alpha,\beta) =&  - ( - 5\,\beta ^{3} + 52\,\alpha ^{2}
\,\beta ^{2} - 35\,\alpha ^{2}\,\beta ^{3} + 26\,\alpha \,\beta ^{3}
+ 26\,\alpha ^{3}\,\beta  - 19\,\alpha \,\beta ^{2} + 6\, \alpha
^{2}\,\beta ^{4}  \\& \mbox{} - 35\,\alpha ^{3}\,\beta ^{2}-
5\,\alpha ^{3} + 6\,\alpha ^{4}\,\beta ^{2} - 19\, \alpha
^{2}\,\beta  + 2\,\alpha ^{4} - 7\,\alpha ^{4}\,\beta  + 12\,\alpha
^{3}\,\beta ^{3} + 2\,\beta ^{4}  \\&- 7\,\alpha \,\beta ^{4})/(
 - \beta  - \alpha  + \alpha \,\beta )
\end{split}
\end{align}
\begin{align}
\begin{split}
{{S}_{5,1}}(\alpha,\beta) =& ( - 133\,\alpha ^{4}\,\beta ^{3} - 89\, \alpha
^{4}\,\beta  - 133\,\alpha ^{3}\,\beta ^{4} + 176\,\alpha
^{2}\,\beta ^{4} - 89\,\alpha \,\beta ^{4} + 176\,\alpha ^{4}\,
\beta ^{2} + 18\,\alpha ^{3}\,\beta ^{5} \\& \mbox{} + 18\,\alpha
^{5}\,\beta ^{3} + 17\,\alpha \,\beta ^{5}
 + 17\,\alpha ^{5}\,\beta  - 35\,\alpha ^{5}\,\beta ^{2} - 35\,
\alpha ^{2}\,\beta ^{5} - 12\,\alpha ^{3} - 319\,\alpha ^{3}\, \beta
^{2} \\& \mbox{} + 136\,\alpha ^{3}\,\beta  - 319\,\alpha
^{2}\,\beta ^{3}
 + 236\,\alpha ^{2}\,\beta ^{2} - 60\,\alpha ^{2}\,\beta  + 136\,
\alpha \,\beta ^{3} - 60\,\alpha \,\beta ^{2} - 12\,\beta ^{3}
 \\&
\mbox{} + 310\,\alpha ^{3}\,\beta ^{3} + 36\,\alpha ^{4}\,\beta ^{4}
+ 10\,\alpha ^{4} + 10\,\beta ^{4})/(2\,( - 1 + \beta )\,(
 - 1 + \alpha )\,( - \beta  - \alpha  + \alpha \,\beta ))
\end{split}
\end{align}
\begin{align}
\begin{split}
{{S}_{5,0}}(\alpha,\beta) =&  - (5\,\alpha ^{4}\,\beta ^{3} + 6\, \alpha
^{4}\,\beta  + 5\,\alpha ^{3}\,\beta ^{4} - 20\,\alpha ^{2 }\,\beta
^{4} + 6\,\alpha \,\beta ^{4} - 40\,\alpha ^{3}\,\beta ^{3} + \alpha
^{5}\,\beta ^{2} - 20\,\alpha ^{4}\,\beta ^{2} \\& \mbox{} -
2\,\alpha \,\beta ^{5} - 2\,\alpha ^{5}\,\beta  + \alpha ^{2}\,\beta
^{5} + 28\,\alpha ^{3}\,\beta ^{2} + 28\, \alpha ^{2}\,\beta ^{3} +
2\,\beta ^{5} + 2\,\alpha ^{5})/(2\, \alpha \,\beta  \\& ( - \beta -
\alpha  + \alpha \,\beta ))
\end{split}
\end{align}
\begin{align}
\begin{split}
{{D}_{5,1}}(\alpha,\beta) =&  - ( - 178\,\alpha ^{4}\,\beta ^{3} + 2 \,\alpha
^{4}\,\beta  - 178\,\alpha ^{3}\,\beta ^{4} + 112\, \alpha
^{2}\,\beta ^{4} + 2\,\alpha \,\beta ^{4} + 112\,\alpha ^{ 4}\,\beta
^{2}  \\& \mbox{} - 28\,\alpha ^{3}\,\beta ^{5}- 28\,\alpha
^{5}\,\beta ^{3} - 33\,\alpha \,\beta ^{5}
 - 33\,\alpha ^{5}\,\beta  + 45\,\alpha ^{5}\,\beta ^{2} + 10\,
\alpha ^{5}\,\beta ^{4} + 10\,\alpha ^{4}\,\beta ^{5}  \\& \mbox{}
+ 45\, \alpha
^{2}\,\beta ^{5}+ 14\,\alpha ^{3}\,\beta ^{6} +
14\,\alpha ^{6}\,\beta ^{ 3} + 13\,\alpha ^{6}\,\beta  + 13\,\alpha
\,\beta ^{6} - 27\, \alpha ^{2}\,\beta ^{6} - 27\,\alpha ^{6}\,\beta
^{2} \\& \mbox{} + 6\,\beta ^{5} + 6\,\alpha ^{5} - 200\,\alpha
^{3}\,\beta ^{2} + 24\,\alpha ^{3}\,\beta
 - 200\,\alpha ^{2}\,\beta ^{3} + 64\,\alpha ^{2}\,\beta ^{2} +
24\,\alpha \,\beta ^{3} \\& \mbox{} + 370\,\alpha ^{3}\,\beta ^{3} + 62\, \alpha
^{4}\,\beta ^{4} - 8\,\alpha ^{4} - 8\,\beta ^{4})/(2\,(
- 1 + \beta )\,(
 - 1 + \alpha )\,( - \beta  - \alpha  + \alpha \,\beta ))
\end{split}
\end{align}
\begin{align}
\begin{split}
{{T}_{5,2}}(\alpha,\beta) =& (18\,\alpha ^{5}\,\beta ^{3} - 24\, \alpha
^{5}\,\beta ^{2} + 5\,\alpha ^{5}\,\beta  + \alpha ^{5} + 36\,\alpha
^{4}\,\beta ^{4} - 84\,\alpha ^{4}\,\beta ^{3} + 78\, \alpha
^{4}\,\beta ^{2} - 25\,\alpha ^{4}\,\beta  \\& \mbox{} - \alpha ^{4}
+ 18\,\alpha ^{3}\,\beta ^{5} + 98\,\alpha ^{3}\,\beta ^{3} -
60\,\alpha ^{3}\,\beta ^{2} + 24\,\alpha ^{3} \,\beta  - 84\,\alpha
^{3}\,\beta ^{4} - 24\,\alpha ^{2}\,\beta ^{5} \\& + 78\,\alpha
^{2}\,\beta ^{4} \mbox{} + 2\,\alpha ^{2}\,\beta ^{2} -
60\,\alpha ^{2}\,\beta ^{3 } + 5\,\alpha \,\beta ^{5} - 25\,\alpha
\,\beta ^{4} + 24\,\alpha
 \,\beta ^{3}\\& + \beta ^{5} - \beta ^{4})/(2\,\alpha \,\beta \,(
 - 1 + \alpha )
( - 1 + \beta ))
\end{split}
\end{align}
\begin{align}
\begin{split}
{{T}_{5,1}}(\alpha,\beta) =&  - (12\,\alpha ^{5}\,\beta ^{3} - 20\, \alpha
^{5}\,\beta ^{2} + 9\,\alpha ^{5}\,\beta  - \alpha ^{5} + 24\,\alpha
^{4}\,\beta ^{4} - 88\,\alpha ^{4}\,\beta ^{3} + 94\, \alpha
^{4}\,\beta ^{2} \\& - 31\,\alpha ^{4}\,\beta  \mbox{} + \alpha ^{4}
+ 12\,\alpha ^{3}\,\beta ^{5} + 178\,\alpha
 ^{3}\,\beta ^{3} - 124\,\alpha ^{3}\,\beta ^{2} + 24\,\alpha ^{3
}\,\beta  - 88\,\alpha ^{3}\,\beta ^{4} \\& \mbox{}- 20\,\alpha ^{2}\,\beta
^{5}  + 94\,\alpha ^{2}\,\beta ^{4} + 46\,\alpha
^{2}\,\beta ^{ 2} - 124\,\alpha ^{2}\,\beta ^{3} + 9\,\alpha \,\beta
^{5} - 31\, \alpha \,\beta ^{4} \\&+ 24\,\alpha \,\beta ^{3} - \beta
^{5} + \beta ^{4})/(\alpha \,\beta   ( - 1 + \alpha )\,( - 1 +
\beta ))
\end{split}
\end{align}
\begin{align}
\begin{split}
{{T}_{5,0}}(\alpha,\beta) =&  - (7\,\beta ^{4} + 6\,\alpha ^{4}\, \beta ^{2} +
32\,\alpha ^{3}\,\beta  + 6\,\alpha ^{2}\,\beta ^{4}
 + 32\,\alpha \,\beta ^{3} + 12\,\alpha ^{3}\,\beta ^{3} - 14\,
\alpha ^{4}\,\beta  + 7\,\alpha ^{4} \\& \mbox{} + 66\,\alpha
^{2}\,\beta ^{2} - 70\,\alpha ^{3}\,\beta ^{ 2} - 14\,\alpha \,\beta
^{4} - 70\,\alpha ^{2}\,\beta ^{3})/(2\, \alpha \,\beta )
\end{split}
\end{align}

\section{Real--emission coefficients at NLO and NNLO\label{sec:real_explicit}}

Below we list the real--emission coefficients
$c_i^{(n)}(\alpha,\beta=r\alpha)$ in \eq{Ri_borel} at NLO ($n=1$)
and NNLO ($n=2$) of each of the five structure functions, $i=1$ to
5. Note that these expressions include a singular (non-integrable)
piece for $r\to 0$. According to the default subtraction
prescription we use, \eq{plus}, the plus prescription is defined
with respect to $r=\beta/\alpha$. The coefficients
$c_i^{(n)}(\alpha,\beta=r\alpha)$ are therefore understood to be
separated as in~\eq{separation}, where the the singular part that is
put under the plus prescription as given in~\eq{ci_12_sing}.
The NLO coefficients are:
\begin{align}
\begin{split}
c_1^{(1)} (\alpha,\alpha r)&= \frac{\left(2 \alpha^2-10
\alpha+7\right) r^3+\left(2 \alpha^2-4 \alpha-7\right) r^2-(10
\alpha+7) r+7}{4 (r-1)
   r}\\
&-\frac{\left(r^4+r^2\right) \alpha^2-2 r \left(r^3+2 r^2+2 r+1\right) \alpha+2
   \left(r^2-1\right)^2 }{2 (r-1)^2 r} \ln r\end{split}
\end{align}
\begin{align}
\begin{split}
c_2^{(1)}(\alpha, \alpha r) &= \frac{\left(-6 \alpha^3+20
\alpha^2-21 \alpha+7\right) r^3-
\left(6 \alpha^3-56 \alpha^2+57 \alpha-10\right) r^2}
{\alpha (r-1)^3 }\\
&+\frac{\left(20   \alpha^2-57 \alpha+38\right) r}{\alpha (r-1)^3}
+\frac{(10-21 \alpha) r+7}{\alpha (r-1)^3 r} \\&+
2\left[\frac{\left(\alpha^3-3 \alpha^2+4
 \alpha-2\right) r^4
 +\left(4 \alpha^3-21 \alpha^2+18 \alpha-2\right) r^3 }
 {\alpha (r-1)^4 }
\right.
\\& \left.+\frac{
\left(\alpha^3-21 \alpha^2+34 \alpha-14\right)
   r^3+\left(-3 \alpha^2+18 \alpha-14\right) r^2+(4 \alpha-2) r-2}{\alpha (r-1)^4 r}
 \right]\ln r
\end{split}
\end{align}
\begin{align}
\begin{split}
c_3^{(1)}(\alpha, \alpha r) &= \frac{\left(4 \alpha^2-10
\alpha+7\right) r^2+(2-10 \alpha) r+7}{2 \alpha (r-1) r}
\\&-\frac{\left(\alpha^2-2
   \alpha+2\right) r^3+\left(\alpha^2-6 \alpha+2\right) r^2-2 (\alpha-1) r+2}{\alpha (r-1)^2 r} \, \ln r
\end{split}
\end{align}
\begin{align}
\begin{split}
c_4^{(1)}(\alpha, \alpha r) &=
\frac{22 r^2+28 r-\alpha \left(r^3+5 r^2+5 r+1\right)+22}{\alpha \,(r-1)^3}\\&
+\frac{4 \left((\alpha-2) r^3+(\alpha-7)
   r^2+(\alpha-7) r-2\right)}{\alpha\, (r-1)^4} \ln r
\end{split}
\end{align}
\begin{align}
\begin{split}
c_5^{(1)}(\alpha, \alpha r) &= \frac{\left(-6 \alpha^2+14
\alpha-7\right) r^3-2 \left(6 \alpha^2-35 \alpha+16\right) r^2
-\left(6 \alpha^2-70 \alpha+66\right) r }{2
\alpha (r-1)^3 }
\\&+\frac{2 (7 \alpha-16) r-7}{2
\alpha (r-1)^3 r}+\left[\frac{\left(\alpha^2-2 \alpha+2\right)
   r^4+\left(5 \alpha^2-20 \alpha+6\right) r^3}{\alpha (r-1)^4 }
   \right.
\\&+
\left.
\frac{\left(5 \alpha^2-40 \alpha+28\right) r^2
+\left(\alpha^2-20 \alpha+28\right)
   r}{\alpha (r-1)^4 }+
\frac{-2 (\alpha-3) r+2}{\alpha (r-1)^4 r}  \right]\ln r
\end{split}
\end{align}

The NNLO coefficients in the large--$\beta_0$ limit are:
\begin{align}
\label{c_i_2}
\begin{split}
c_i^{(2)}(\alpha, \alpha r) &= \frac{-1}{\left[\alpha (1-r)\right]^{2 y_i}} \Bigg\{
{\mathbb{A}_i}(\alpha,r) \,\ln(r) + {\mathbb{B}_i}(\alpha,r)\,
\frac{\alpha (1-r) (1+r -\alpha
r)}{r} \, \ln (1+r-\alpha r)
\\
&+{\mathbb{T}_i}(\alpha,r) + 2 \alpha  (1+r-\alpha r) \,S_{i,0}(\alpha,\alpha r) \bigg[
\mathrm{Li}_2\left(\frac{1-\alpha r}{1+r-\alpha r}\right)-
 \mathrm{Li}_2\left(\frac{r \,(1-\alpha)}{1+r-\alpha r}\right)
   \\&  +\frac12 \ln^2(r) -\ln(r)\, \ln (1+r-\alpha r) \bigg] \Bigg\}
 + \left(\frac53 -2\ln(\alpha)-\ln(r)\right) c_i^{(1)}(\alpha, \alpha r)
\end{split}
\end{align}
with
$$\mathbb{T}_i(\alpha,r)= \alpha (1-r)\left[T_{i,1}(\alpha, \alpha r)
+\frac12 T_{i,2}(\alpha, \alpha r)\right]$$
where $T_{i,j}(\alpha, \beta)$ and $S_{i,0}(\alpha,\beta)$ in (\ref{c_i_2})
are given in Appendix~\ref{sec:DST_explicit_expressions} and
$y_i=[1,2,1,2,2]$. The other functions
${\mathbb{A}_i}(\alpha,r)$ and ${\mathbb{B}_i}(\alpha,r)$ are:
\begin{align}
\begin{split}
\mathbb{A}_1(\alpha,r)&=\frac{\alpha^2}{2 r (\alpha r-1)^2}\,
\bigg[ \left(2 \alpha^4-4 \alpha^3+3
\alpha^2\right) r^6+\left(2 \alpha^4-7 \alpha^3+6 \alpha^2-5
\alpha\right) r^5
\\
&+\left(\alpha^4-11 \alpha^3+8
   \alpha^2+5 \alpha+2\right) r^4
   +\left(19 \alpha^2-8 \alpha-6\right) r^3\\&
   +\left(-5 \alpha^2-12 \alpha+4\right) r^2+(6 \alpha+2)
   r-2\bigg]
\end{split}
\end{align}

\begin{align}
\begin{split}
\mathbb{A}_2(\alpha,r)&=\alpha^3\bigg[ \left(-8 \alpha^3+20
\alpha^2-18 \alpha+6\right) r^4+\left(-20 \alpha^3+78 \alpha^2-84
\alpha+26\right) r^3\\& -\left(2 \alpha^3-36
   \alpha^2+76 \alpha-36\right) r^2
+\left(-2 \alpha^2-16 \alpha+20\right) r+8 \alpha-\frac{4}{r}
\bigg]
\end{split}
\end{align}

\begin{align}
\begin{split}
\mathbb{A}_3(\alpha,r)&= \frac{\alpha}{r (\alpha
   r-1)}\bigg[
   \left(3 \alpha^3-4
\alpha^2+3 \alpha\right) r^4+\left(\alpha^3-8 \alpha^2+6
\alpha-2\right) r^3\\&+ \left(\alpha^2+6 \alpha-4\right) r^2-4
\alpha
r+2 \bigg]
\end{split}
\end{align}
\begin{align}
\begin{split}
\mathbb{A}_4(\alpha,r)&=\frac{\alpha^3}{(\alpha r-1)^3}\,
\bigg[-\alpha^4 r^7-\left(10 \alpha^4-21
\alpha^3+4 \alpha^2\right) r^6-\left(13 \alpha^4-64 \alpha^3+68
\alpha^2-6 \alpha\right)
   r^5\\&
+\left(-6 \alpha^4+67 \alpha^3-138 \alpha^2+72 \alpha-2\right) r^4+\left(10 \alpha^3-116 \alpha^2+136 \alpha-26\right)
   r^3\\&+\left(8 \alpha^2+76 \alpha-48\right) r^2+(-20 \alpha-16) r+8
\bigg]
\end{split}
\end{align}

\begin{align}
\begin{split}
\mathbb{A}_5(\alpha,r)&=\frac{\alpha^3}{r\,(\alpha r-1)^2}\,
\bigg[ \left(-4 \alpha^4+6 \alpha^3-3
\alpha^2\right) r^7+\left(-14 \alpha^4+47 \alpha^3-35 \alpha^2+8
\alpha\right) r^6 \\&+\left(-11 \alpha^4+80
   \alpha^3-132 \alpha^2+52 \alpha-4\right) r^5
+\left(-\alpha^4+29 \alpha^3-133 \alpha^2+134 \alpha-26\right) r^4
\\&+\left(-20 \alpha^2+84
   \alpha-42\right) r^3+\left(5 \alpha^2-2 \alpha-18\right) r^2+(4-6 \alpha) r+2
\bigg]
\end{split}
\end{align}

\begin{align}
\begin{split}
\mathbb{B}_1(\alpha,r)&=\frac{\alpha}
{2 (\alpha-1)^2  (\alpha r-1)^2}\,
\bigg[ \left(-\alpha^5+7 \alpha^4-11
\alpha^3+5 \alpha^2\right) r^4\\&+\left(-\alpha^5+4 \alpha^4-10
\alpha^3+17 \alpha^2-9 \alpha\right) r^3 +\left(7
   \alpha^4-10 \alpha^3-6 \alpha^2+3 \alpha+4\right) r^2\\&
   +\left(-11 \alpha^3+17 \alpha^2+3 \alpha-8\right) r+5 \alpha^2-9 \alpha+4
\bigg]
\end{split}
\end{align}

\begin{align}
\begin{split}
\mathbb{B}_2(\alpha,r)&=2 \alpha^2 \,\big[ \left(3 \alpha^2-8
\alpha+5\right) r^3+\left(3 \alpha^2-26 \alpha+13\right) r^2+(13-8
\alpha) r+5 \big]
\end{split}
\end{align}

\begin{align}
\begin{split}
\mathbb{B}_3(\alpha,r)&=    \frac{ \left(-2 \alpha^3+6 \alpha^2-5
\alpha\right) r^2+ \left(6 \alpha^2-8 \alpha+4\right) r-5 \alpha+4
}{(\alpha-1)  (\alpha r-1)}
\end{split}
\end{align}

\begin{align}
\begin{split}
\mathbb{B}_4(\alpha,r)&= \frac{\alpha^2}{(\alpha-1)^3  (\alpha r-1)^3}\,
\bigg[\left(\alpha^6-2
\alpha^5+\alpha^4\right) r^6+\left(5 \alpha^6-38 \alpha^5+68
\alpha^4-37 \alpha^3+4 \alpha^2\right) r^5 \\&
+\left(5 \alpha^6-52
   \alpha^5+171 \alpha^4-230 \alpha^3+104 \alpha^2-6 \alpha\right) r^4
   \\&+\left(\alpha^6-38 \alpha^5+171 \alpha^4-306 \alpha^3+282 \alpha^2-100
   \alpha+2\right) r^3\\&
-\left(2 \alpha^5-68 \alpha^4+230 \alpha^3-282 \alpha^2+160
\alpha-34\right) r^2\\&+\left(\alpha^4-37 \alpha^3+104
   \alpha^2-100 \alpha+34\right) r+4 \alpha^2
-6 \alpha+2   \bigg]
\end{split}
\end{align}

\begin{align}
\begin{split}
\mathbb{B}_5(\alpha,r)&=
\frac{\alpha^2}{(\alpha-1)^2  (\alpha r-1)^2}\,
\bigg[(\alpha-1)^2 \alpha^2 (3 \alpha-5)
r^5\\&+\alpha \left(6 \alpha^4-49 \alpha^3+87 \alpha^2-55
\alpha+12\right) r^4
\\&
+\left(3 \alpha^5-49 \alpha^4+160
   \alpha^3-180 \alpha^2+68 \alpha-6\right) r^3\\&
   +\left(-11 \alpha^4+87 \alpha^3-180 \alpha^2+140 \alpha-30\right) r^2
\\&+\left(13
   \alpha^3-55 \alpha^2+68 \alpha-30\right) r-5 \alpha^2+12
   \alpha-6
   \bigg]
\end{split}
\end{align}

\section{Virtual coefficients at NLO and NNLO\label{sec:Virtual_NLO_NNLO}}
The coefficients $v_i^{(1,2)}(\alpha)$ entering (\ref{Vi_reg_borel})
for the five structure functions, $i=1$ to $5$ are listed below.
These coefficients are computed by expanding \eq{BV_reg_expl}.
Let us recall that these expressions correspond to the
infrared--subtraction procedure detailed in Sec.~\ref{sec:Virt}, where
$\left.B[V_0](\alpha,u)\right\vert_{\sing}$ is defined according to
(\ref{virt_sing}), where the plus prescription is defined with
respect to $r=\beta/\alpha$, as in (\ref{plus}).

Let us also recall that according to \eq{BV_reg_expl}
there is a simple all--order relation between the virtual corrections
for the structure functions $i=3$ and $i=1$, namely
\begin{align}
\begin{split}
v_3^{(n)}(\alpha) &=  \frac{2}{\alpha}\, v_1^{(n)}(\alpha)\,.
\end{split}
\end{align}
The NLO and NNLO coefficients for the other structure functions are:
\begin{align}
\begin{split}
v_1^{(1)}(\alpha) &=  \frac{\alpha}{2}\bigg[- 2\,\mathrm{Li}_2(1 - \alpha ) -
{\displaystyle \frac {( - 3 + 2\,\alpha )\,\mathrm{ln}(\alpha )}{
 - 1 + \alpha }}  - {\displaystyle \frac {2\,\pi ^{2}}{3}}  -
{\displaystyle \frac {5}{2}}\bigg]
\end{split}
\end{align}
\begin{align}
\begin{split}
v_1^{(2)}(\alpha) &=\frac{\alpha}{2}\bigg[
\left(\frac{2 \alpha-3}{\alpha-1}-2 \ln (1-\alpha)\right) \ln ^2(\alpha)+\left(\frac{33-14 \alpha}{6 (\alpha-1)}+2 \pi
   ^2\right) \ln (\alpha)
\\&+\frac{(16-19\, \alpha)\, \text{Li}_2(1-\alpha)}{3 (\alpha-1)}-2\, \text{Li}_3(1-\alpha)-4
  \, \text{Li}_3(\alpha)+2 \zeta_3-\frac{79\, \pi ^2}{36}-\frac{71}{12}\bigg]
\end{split}
\end{align}
\begin{align}
\begin{split}
v_2^{(1)}(\alpha) &=  - 4\,\mathrm{ln}(\alpha ) - 4\,\mathrm{
Li}_2(1 - \alpha ) - {\displaystyle \frac {4\,\pi ^{2}}{3 }} - 5
\end{split}
\end{align}
\begin{align}
\begin{split}
v_2^{(2)}(\alpha) &=
(4-4 \ln (1-\alpha)) \ln^2(\alpha)+\left(4 \pi ^2-\frac{2 \, (7 \alpha-4)}{3 (\alpha-1)}\right)
\ln   (\alpha)\\&
-\frac{38}{3} \text{Li}_2(1-\alpha)-4 \,\text{Li}_3(1-\alpha)-8  \,\text{Li}_3(\alpha)+4 \zeta_3
-\frac{79 \pi ^2}{18}-\frac{71}{6}
\end{split}
\end{align}
\begin{align}
\begin{split}
v_4^{(1)}(\alpha) &=  - {\displaystyle \frac {2\,( - 1 + 2\, \alpha
)\,\mathrm{ln}(\alpha )}{( - 1 + \alpha )^{2}}}  + {\displaystyle
\frac {2}{ - 1 + \alpha }}
\end{split}
\end{align}
\begin{align}
\begin{split}
v_4^{(2)}(\alpha) &= {\displaystyle \frac {2\,( - 1 + 2\,\alpha
 )\,\mathrm{Li}_2(1 - \alpha )}{( - 1 + \alpha )^{2}}}
 + {\displaystyle \frac {2\,( - 1 + 2\,\alpha )\,\mathrm{ln}(
\alpha )^{2}}{( - 1 + \alpha )^{2}}}  \\ & \mbox{} - {\displaystyle
\frac {1}{3}} \,{\displaystyle \frac {(
 - 25 + 38\,\alpha )\,\mathrm{ln}(\alpha )}{( - 1 + \alpha )^{2}}
}  + {\displaystyle \frac {19}{3\,( - 1 + \alpha )}}
\end{split}
\end{align}
\begin{align}
\begin{split}
v_5^{(1)}(\alpha) &= 2\,\mathrm{Li}_2(1 - \alpha ) + {\displaystyle
\frac {(2\,\alpha ^{2} - 2\,\alpha  + 1)\,\mathrm{ ln}(\alpha )}{( -
1 + \alpha )^{2}}}  + {\displaystyle \frac {2\, \pi ^{2}}{3}}  +
{\displaystyle \frac {5\,\alpha  - 7}{2\,( - 1
 + \alpha )}}
\end{split}
\end{align}
\begin{align}
\begin{split}
v_5^{(2)}(\alpha) &=
\left(2 \ln (1-\alpha)-\frac{2 \,(\alpha-1) \alpha+1}{(\alpha-1)^2}\right) \ln ^2(\alpha)+
\left(\frac{14 \alpha^2+16
   \alpha-17}{6 (\alpha-1)^2}-2 \pi ^2\right) \ln (\alpha)-2 \zeta_3
\\&+\frac{22+\alpha \,(19 \alpha-44)}{3
   (\alpha-1)^2}\, \text{Li}_2(1-\alpha)+2 \,\text{Li}_3(1-\alpha)+4 \,\text{Li}_3(\alpha)+
\frac{109-71 \alpha}{12-12 \alpha}+\frac{79 \pi ^2}{36}
\end{split}
\end{align}

\section{Polynomials entering the triple differential width\label{sec:Polylomials_NNLO}}

Below we list the polynomials in $r=\beta/\alpha$, $\alpha$ and $x$ entering the triple
differential width summarized in Sec.~\ref{sec:put_together}.
The polynomials entering the NLO coefficient through \eq{NLO_result_reg} are:
\begin{align}
\begin{split}
{\mathbb Q}_1(\alpha,\, r,\, x)&= \alpha ^{2}\,(
  1 - \alpha )\,(\alpha ^{2} - 2\,\alpha  + 2)\,r^{6}+ \alpha \,(  1 - \alpha )\,\times
 \\ & \hspace*{-40pt}
\mbox{} ( - 7\,\alpha ^{2} + 2\,x\,
\alpha ^{2} + 5\,\alpha  - 4\,x\,\alpha  - 6 + 4\,x)\,r^{5} - (
 - 10\,x\,\alpha  + 42\,\alpha ^{3} - 2\,x^{2} + x^{2}\,\alpha ^{3}\\ & \hspace*{-40pt}
\mbox{}  + 29\,x\,\alpha ^{2} - 4 + 4\,\alpha
 \,x^{2} + 4\,x\,\alpha ^{4} - 3\,x^{2}\,\alpha ^{2} + 4\,\alpha
^{5} - 20\,\alpha ^{4} + 6\,x + 3\,\alpha- 28\,x\,\alpha ^{3}  \\ & \hspace*{-40pt}
\mbox{}  - 25\,\alpha ^{2})r^{4}\mbox{} - (
 - 64\,x\,\alpha  + 4 - 98\,\alpha ^{2} + 4\,x\,\alpha ^{4} - 21
\,x^{2}\,\alpha ^{2} + 46\,\alpha  - 40\,x\,\alpha ^{3} \\ & \hspace*{-40pt}
\mbox{} - 4\,x - 20\,\alpha ^{4} + 95\,x\,\alpha ^{2} + 18\,
\alpha \,x^{2} + 68\,\alpha ^{3} + 4\,x^{2}\,\alpha ^{3})r^{3}
\mbox{} - (75\,x\,\alpha ^{2} - 80\,x\,\alpha  \\ & \hspace*{-40pt}
\mbox{} + 34\,\alpha \,x^{2} + 32\,\alpha ^{3} + 52\,x + \alpha
^{5} - 21\,x^{2}\,\alpha ^{2} + x^{2}\,\alpha ^{3} - 22\,x^{2} +
52\,\alpha  + 2\,x\,\alpha ^{4} - 7\,\alpha ^{4} \\ & \hspace*{-40pt}
\mbox{} - 48\,\alpha ^{2} - 30 - 28\,x\,\alpha ^{3})r^{2}\mbox{}
 + ( - 16\,\alpha  - 18\,\alpha \,x^{2} + 2\,x^{2} + 3\,x^{2}\,
\alpha ^{2} - 10 - 39\,x\,\alpha ^{2} \\ & \hspace*{-40pt}
\mbox{} - 20\,\alpha ^{3} + 43\,\alpha ^{2} + 8\,x + 6\,x\,\alpha
 ^{3} + 44\,x\,\alpha  + 3\,\alpha ^{4})r\mbox{} - 2\,\alpha ^{3}
 - (6\,x - 13)\,\alpha ^{2} \\ & \hspace*{-40pt}
\mbox{} - (27 + 4\,x^{2} - 26\,x)\,\alpha  + 16 + 10\,x^{2} - 26
\,x
\end{split}
\end{align}
\begin{align}
\begin{split}
{\mathbb Q}_2(\alpha,\, r,\, x)&=2\,\alpha ^{2}\,r^{2} - 10\,\alpha \,r^{2} + 7\,r^{2} + 2
\,\alpha ^{2}\,r - 4\,\alpha \,r - 7\,r - 10\,\alpha
\end{split}
\end{align}
\begin{align}
\begin{split}
{\mathbb Q}_3(\alpha,\, r,\, x)&=
\left(6\,\alpha ^{3} - 20\,\alpha ^{2} + 21\,\alpha  - 7\right)r^3
 +\left(- 17 - 56\,\alpha ^{2} + 57\,\alpha
 + 6\,\alpha ^{3}\right) r^{2} \\ & +\left(- 20\,\alpha ^{2}
 - 17 + 57\,\alpha \right)r + 21\,\alpha  - 31
 \end{split}
\end{align}

The polynomials entering the NNLO coefficient through \eq{NNLO_result_reg} are:
\begin{align}
\begin{split}
{\mathbb{P}_{1}(\alpha,\, r,\,x)} &=   \alpha ^{3}\,(  1 - \alpha )\,(3\,
\alpha ^{2} - 12\,\alpha  + 10)\,r^{6} \\ & \hspace*{-20pt}
\mbox{} - \alpha ^{2}\,(38 + 3\,\alpha ^{4} - 20\,x - 38\,x\,
\alpha ^{2} + 48\,x\,\alpha  - 22\,\alpha ^{3} - 83\,\alpha  + 63
\,\alpha ^{2} + 10\,x\,\alpha ^{3})\,r^{5}   \\ & \hspace*{-20pt}
-\alpha (26\,\alpha \,x^{2} - 46 + 3\,\alpha ^{5} + 6\,x^{2}\,
\alpha ^{3} + 4\,x\,\alpha ^{4} - 10\,x^{2} - 58\,\alpha ^{4} -
140\,x\,\alpha  - 82\,x\,\alpha ^{3} \\ & \hspace*{-20pt}
\mbox{} - 160\,\alpha ^{2} - 22\,x^{2}\,\alpha ^{2} + 170\,x\,
\alpha ^{2} + 155\,\alpha ^{3} + 110\,\alpha  + 48\,x)r^{4}
\mbox{} - ( - 334\,\alpha ^{3} - 28\,x \\ & \hspace*{-20pt}
\mbox{} + 155\,\alpha ^{4} + 10\,x^{2} - 22\,\alpha ^{5} + 136\,x
\,\alpha  - 68\,\alpha  - 64\,x^{2}\,\alpha ^{3} + 242\,\alpha ^{
2} - 52\,\alpha \,x^{2} \\ & \hspace*{-20pt}
\mbox{} + 284\,x\,\alpha ^{3} - 82\,x\,\alpha ^{4} - 320\,x\,
\alpha ^{2} + 6\,x^{2}\,\alpha ^{4} + 100\,x^{2}\,\alpha ^{2} + 3
\,\alpha ^{6} + 10\,x\,\alpha ^{5} + 18)r^{3}   \\ & \hspace*{-20pt}
+ (15\,\alpha ^{5} + 22\,x^{2}\,\alpha ^{3} - 63\,\alpha ^{4} + 320
\,x\,\alpha ^{2} - 26\,x^{2} + 44\,x + 104\,\alpha \,x^{2} + 38\,
x\,\alpha ^{4} \\ & \hspace*{-20pt}
\mbox{} - 100\,x^{2}\,\alpha ^{2} + 160\,\alpha ^{3} - 232\,x\,
\alpha  + 144\,\alpha  - 242\,\alpha ^{2} - 18 - 170\,x\,\alpha
^{3})r^{2}\mbox{} - (136\,x\,\alpha  \\ & \hspace*{-20pt}
\mbox{} - 140\,x\,\alpha ^{2} + 26\,x^{2} + 26\,x^{2}\,\alpha ^{2
} - 44\,x + 18 + 22\,\alpha ^{4} + 110\,\alpha ^{2} - 52\,\alpha
\,x^{2} - 68\,\alpha  \\ & \hspace*{-20pt}
\mbox{} - 83\,\alpha ^{3} + 48\,x\,\alpha ^{3})r\mbox{} + 10\,
\alpha ^{3} + 2\,( - 19 + 10\,x)\,\alpha ^{2} + 2\,(23 - 24\,x +
5\,x^{2})\,\alpha  - 18 \\ & \hspace*{-20pt}
\mbox{} + 28\,x - 10\,x^{2}
\end{split}
\end{align}

\begin{align}
\begin{split}
{\mathbb{P}_{2}(\alpha,\, r,\,x)} &= -2\,\alpha ^{3}\,(  1 - \alpha )\,(34\,
\alpha ^{2} - 74\,\alpha  + 59)\,r^{7} \\ & \hspace*{-20pt}
\mbox{} - \alpha^{2}\,( 1 - \alpha )\,(24\,\alpha ^{3} - 424\,
\alpha ^{2} + 184\,x\,\alpha ^{2} + 557\,\alpha  - 344\,x\,\alpha
  - 460 + 236\,x)\,r^{6}  \\ & \hspace*{-20pt}
+ 2 \alpha (289\,x + 650\,\alpha  - 283 - 1204\,\alpha ^{2} + 1229\,x
\,\alpha ^{2} + 1291\,\alpha ^{3} - 822\,x\,\alpha  + 52\,x^{2}\,
\alpha ^{3} \\ & \hspace*{-20pt}
\mbox{} - 804\,x\,\alpha ^{3} - 539\,\alpha ^{4} + 157\,\alpha \,
x^{2} + 76\,x\,\alpha ^{4} - 59\,x^{2} - 150\,x^{2}\,\alpha ^{2}
 + 85\,\alpha ^{5})r^{5}\mbox{}  \\ & \hspace*{-20pt}
+ (3320\,\alpha ^{4} - 1208\,x^{2}\,\alpha ^{3} + 224 - 667\,\alpha
 + 200\,x^{2}\,\alpha ^{4} - 5680\,\alpha ^{3} - 342\,x - 4674\,x
\,\alpha ^{2} \\ & \hspace*{-20pt}
\mbox{} + 5352\,x\,\alpha ^{3} + 118\,x^{2} - 554\,\alpha \,x^{2}
 + 3451\,\alpha ^{2} + 1416\,x\,\alpha  - 1732\,x\,\alpha ^{4} +
24\,\alpha ^{6} \\ & \hspace*{-20pt}
\mbox{} + 260\,x\,\alpha ^{5} + 1380\,x^{2}\,\alpha ^{2} - 672\,
\alpha ^{5})r^{4}\mbox{} + 2(2592\,\alpha ^{2} + 56 + 2\,x\,
\alpha ^{5} - 127\,\alpha ^{5} \\ & \hspace*{-20pt}
\mbox{} + 1936\,x\,\alpha ^{3} + 727\,\alpha ^{4} - 470\,x\,
\alpha ^{4} + 1120\,x^{2}\,\alpha ^{2} - 3424\,x\,\alpha ^{2} - 2
\,x^{2}\,\alpha ^{4} - 2105\,\alpha ^{3} \\ & \hspace*{-20pt}
\mbox{} - 176\,x + 7\,\alpha ^{6} + 120\,x^{2} - 1150\,\alpha  +
1910\,x\,\alpha  - 820\,\alpha \,x^{2} - 310\,x^{2}\,\alpha ^{3})
r^{3}\mbox{} + (16\,\alpha ^{5} \\ & \hspace*{-20pt}
\mbox{} - 1076\,x + 76\,x^{2}\,\alpha ^{3} - 1524\,\alpha \,x^{2}
 + 560\,x^{2} + 660\,x^{2}\,\alpha ^{2} - 1491\,\alpha ^{3} + 104
\,x\,\alpha ^{4} \\ & \hspace*{-20pt}
\mbox{} - 2178\,\alpha  + 944\,x\,\alpha ^{3} + 2834\,\alpha ^{2}
 - 3424\,x\,\alpha ^{2} + 303\,\alpha ^{4} + 516 + 3744\,x\,
\alpha )r^{2}\mbox{} \\ & \hspace*{-20pt}
- 2( 412\,x + 37\,\alpha ^{4} + 79\,\alpha \,x^{2} + 83\,x^{2}\,\alpha
 ^{2} - 218 - 194\,x^{2} - 16\,x\,\alpha ^{2} + 123\,x\,\alpha ^{
3} - 493\,x\,\alpha  \\ & \hspace*{-20pt}
\mbox{} + 401\,\alpha  - 184\,\alpha ^{2} - 36\,\alpha ^{3})r
\mbox{} + 44\,\alpha ^{3} + ( - 145 + 138\,x)\,\alpha ^{2} \\ & \hspace*{-20pt}
\mbox{} + ( - 296\,x + 94\,x^{2} + 165)\,\alpha  - 64 - 82\,x^{2}
 + 146\,x
\end{split}
\end{align}
\begin{align}
\begin{split}
{\mathbb{P}_{3}(\alpha,\, r,\,x)} &=   \alpha \,(  1 - \alpha )\,(50\,\alpha
^{2} - 142\,\alpha  + 85)\,r^{3} - (50\,\alpha ^{4} + 219\,\alpha
 ^{2} - 142\,\alpha  + 85 - 200\,\alpha ^{3})\,r^{2} \\ & \hspace*{-20pt}
\mbox{} + ( - 304\,\alpha ^{2} + 192\,\alpha ^{3}  + 39\,\alpha  + 85)\,r
 - 142\,( - 1 + \alpha )\,\alpha
\end{split}
\end{align}
\begin{align}
\begin{split}
{\mathbb{P}_{4}(\alpha,\, r,\,x)} &=   (  1 - \alpha )\,(150\,\alpha ^{2} -
218\,\alpha  + 85)\,r^{3}  - ( - 1016\,\alpha ^{2}  + 1035\,\alpha  -
287 + 150\,\alpha ^{3})\,r^{2} \\ & \hspace*{-20pt}
\mbox{}
 + (368\,\alpha ^{2}  - 1035\,\alpha  + 395)
\,r + 457 - 303\,\alpha
\end{split}
\end{align}
\begin{equation}
{\mathbb{P}_{5}(\alpha,\, r,\,x)} =  - (85 + 88\,\alpha ^{2} - 142\,\alpha )\,
r +  142\,\alpha  - 123
\end{equation}

\bibliography{database}
\bibliographystyle{JHEP}

\end{document}